\documentclass[acmsmall]{acmart}

\AtBeginDocument{%
  \providecommand\BibTeX{{%
    \normalfont B\kern-0.5em{\scshape i\kern-0.25em b}\kern-0.8em\TeX}}}

\acmYear{2021}
\acmMonth{11}

\newif\ifworkinprogress
\workinprogresstrue

\ifworkinprogress
  \newcommand{\tn}[1]{\textcolor{black}{#1}}
  
  \newcommand{\sr}[1]{\textcolor{black}{#1}}
  \newcommand{\todo}[1]{\textcolor{red}{TODO: #1}}
\else
  \newcommand{\tn}[1]{#1}
  
  \newcommand{\sr}[1]{#1}
  \newcommand{\todo}[1]{\textcolor{red}{[REMOVE] #1}}
\fi

\usepackage{graphicx}
\usepackage{longtable}
\usepackage{hyperref}

\setcopyright{none}

\begin{document}

%\title{Hybrid Collaboration and Meetings: A Survey of Pre-COVID-19 Research and Directions for the Future of Hybrid Work}

\title{What was Hybrid? A Systematic Review of Hybrid Collaboration and Meetings Research}

\author{Thomas Neumayr}
\authornote{Both main authors contributed equally to this research.}
\email{thomas.neumayr@fh-hagenberg.at}
\orcid{0000-0003-3607-8873}
%\email{webmaster@marysville-ohio.com}
\affiliation{%
  \institution{University of Applied Sciences Upper Austria}
  \streetaddress{Softwarepark 11}
  \city{Hagenberg}
  \country{Austria}
  \postcode{4232}
}
\affiliation{%
  \institution{Johannes Kepler University Linz}
  \streetaddress{Altenbergerstraße 69}
  \city{Linz}
  \country{Austria}
  \postcode{4040}
}

\author{Banu Saat\c{c}\.{i}}
\email{banu.saatci@cc.au.dk}
\orcid{0000-0002-3366-2356}
\authornotemark[1]
\affiliation{%
  \institution{Aarhus University}
  \streetaddress{Helsingforsgade 14}
  \city{Aarhus}
  \country{Denmark}
  \postcode{8200}}

\author{Sean Rintel}
\email{serintel@microsoft.com}
\orcid{0000-0003-0840-0546}
\affiliation{%
  \institution{Microsoft Research Cambridge}
  \streetaddress{21 Station Rd}
  \city{Cambridge}
  \country{UK}
  \postcode{CB1 2FB}}

\author{Clemens Nylandsted Klokmose}
\email{clemens@cavi.au.dk}
\orcid{0000-0002-1866-0619}
\affiliation{%
  \institution{Aarhus University}
  \streetaddress{Helsingforsgade 14}
  \city{Aarhus}
  \country{Denmark}
  \postcode{8200}}

\author{Mirjam Augstein}
\email{mirjam.augstein@fh-hagenberg.at}
\orcid{0000-0002-7901-3765}
\affiliation{%
  \institution{University of Applied Sciences Upper Austria}
  \streetaddress{Softwarepark 11}
  \city{Hagenberg}
  \country{Austria}
  \postcode{4232}
}

%\renewcommand{\shortauthors}{Neumayr and Saat\c{c}i}

\begin{abstract}
    Interest in hybrid collaboration and meetings (HCM), where several co-located participants engage in coordinated work with remote participants, is gaining unprecedented momentum after the rapid shift in working from home due to the COVID-19 pandemic. However, while the interest is new, researchers have been exploring HCM phenomena for decades, albeit dispersed across diverse research traditions, using different terms, definitions, and frameworks. In this article, we present a systematic literature review of the contexts and tools of HCM in the ACM Digital Library. We obtained approximately 1,200 results, which were narrowed down to 62 key articles. We report on the terms, citations, venues, authors, domains, study types, and data of these publications and present a taxonomic overview based on their reported hybrid settings' actual characteristics. We discuss why the SLR resulted in a relatively small number of publications, and then as a corollary, discuss how some excluded high-profile publications flesh out the SLR findings to provide important additional concepts. The SLR itself covers the ACM until November 2019, so our discussion also includes relevant 2020 and 2021 publications. The end result is a baseline that researchers and designers can use in shaping the post-COVID-19 future of HCM systems.
    
    % Interest in hybrid collaboration and meetings (HCM), where several co-located participants engage in coordinated work with one or more remote participants, is gaining unprecedented momentum after the rapid shift in working from home due to the COVID-19 pandemic. However, while the interest is new, researchers have been exploring HCM phenomena for decades, albeit dispersed across diverse research traditions, using different terms, definitions, and frameworks. In this article, we present a systematic literature review of the contexts and tools of HCM in the ACM Digital Library. We used 72 keywords to obtain approximately 1,200 results, which were narrowed down to 62 key articles. We report on the terms, citations, venues, authors, domains, study types, and data of these publications and present a taxonomic overview based on their reported hybrid settings’ group size, number of locations, levels of synchronicity, software tools, and hardware devices. We discuss why the SLR resulted in a relatively small number of publications, and then as a corollary, discuss how some excluded high-profile publications flesh out the SLR findings to provide important additional concepts. The SLR itself covers the ACM until November 2019, so our discussion also includes relevant 2020 and 2021 publications. The end result is a baseline that researchers and designers can use in shaping the post-COVID-19 future of HCM systems.
    
\end{abstract}




\maketitle

%\textcolor{purple}{Banu's texts}
%\textcolor{blue}{Thomas' texts}

\section{Introduction}
%\textcolor{brown}{Thoughts: Definitions Hybrid Meetings \& Hybrid Collaboration, Significance of the study, the need for the research, confusion of terms that are used interchangeably, problems with everyone connected virtually (remote) it works, but as soon as they get back to hybrid the problems start. Structure of the paper. \textbf{Suggestion: Show stock exchange value of Zoom (Nasdaq: ZM) during the pandemic (from 68.72 USD on January 2, 2020, increasing up to 457.69 USD at its interim peak on September 1, 2020)} \textbf{Suggestion: COVID-19 pandemic -> number of employees working from home in the US has risen from 15\% occasionally working from home to 50\% working from home full-time according to D. Sull, C. Sull, and J. Bersin, ``Five ways leaders can support remote work'', MIT Sloan Management Review, vol. 61, no. 4, pp. 1–10, 2020 (cit. on p. 1).} Many companies we are now talking to tell as that they don't want to go back to working fully from the office but want to make stronger use of WFH.}

\sr{The increasing number of global companies and mobility of workers has paved the path for the rise of partially distributed teams in the modern workplace~\cite{olson_working_2013}. The shift from physical to digital accelerated in the 2020-2021 COVID-19 pandemic, as the need to limit virus transmission forced many organizations to adapt to partial or completely digital ways to do their work~\cite{teevan2021}. Many companies believe that the post-pandemic future of work will be extensively hybrid. For example, drawing on over 31,092 full-time employed or self-employed workers across 31 markets, Microsoft's 2021 Work Trend Index Annual Report finds that 66\% of leaders say that their company is considering redesigning office space for hybrid work and 73\% of remote employees want flexible remote work options to stay \cite{WorkplaceIndexReport2021}.}

%Recently published reports (see below) on remote work as well as the usage of videoconferencing software are apparent indicators of this. With the ``work from home'' policies by companies, 

%The percentage of remote workers in the U.S. has doubled in two weeks after mid-March 2020\footnote{U.S. Workers Discovering Affinity for Remote Work (Gallup Panel 2020) \url{https://news.gallup.com/poll/306695/workers-discovering-affinity-remote-work.aspx}, last access February 17, 2021.}.}
%Online meetings have been at the centre of changes to work 

%It now appears that remote work has turned from being a short-term necessity to a potential long-term opportunity for workers in diverse sectors.

%While it is still too early to discuss the long-term effects of the pandemic on the future of work, we can argue Workers and workplaces are getting more and more aware of and prepared for hybrid forms of collaboration and meetings more than ever, which will eventually lead to further research in this topic. 

\begin{comment} 

Meetings are at the heart of coordinating much of modern work, and naturally the effect of the pandemic was to increase online meetings exponentially. In the first month of mandatory working from home in March 2020, Microsoft reported that total video calls in its Microsoft Teams service grew by over 1000 percent in March 2020 alone~\cite{remoteworktrend}. 
It is not surprising, then, that as we move into a new era of hybrid work, attention is turning to its challenges and opportunities.

\end{comment}

Coordinated work has been a focus of HCI and CSCW research for at least forty years, with a special focus on geodistributed collaboration and meetings. As such, it is natural to turn to this research to draw lessons for a future of hybrid work. However, the popularity of the adjective `hybrid' as a shorthand to refer to distributed combinations of cohorts doing work is itself a pandemic-based phenomenon. Prior to the pandemic, researchers used a variety of terms to refer to geodistributed collaboration and meetings. Adjectives such as `distributed', `online', `virtual', `remote', and `mediated' have been applied to descriptions of activities that span the very general, such as `collaboration', to specific, such as `brainstorming'. These terms speak to different and evolving conceptual standpoints, but have either been used interchangeably, or certain terms have been favoured in research lineages but not been connected to others. This is especially the case for research on hybridity, where the essential elements of co-located participants engaging in coordinated work with one or more remote participants are sometimes specified but often assumed, not conceived of, or accidentally excluded. In sum, the emergent concepts on what we might want to call hybrid collaboration and meetings do not reflect a common understanding, conceptualization, and definition of phenomena that the world is now desperately racing to understand.

% . As the tools %for videoconferencing and remote collaboration have evolved and 
%companies have %become more 
%globalized, the phenomena 
% of remoteness has 
%have diversified %across different types of meetings and collaboration as well, creating the need for new terms to describe the action% in meetings and meeting styles
%These emergent %forms and terms %of remoteness  do not necessarily reflect a common understanding, conceptualization, and definition of the phenomena. %which have caused many 
%Researchers in HCI and CSCW (as well as in other fields) to use the terms above interchangeably% without an acknowledged differentiation

%The inevitable result of technological and workplace evolution combined with the range of disciplinary interests is that the literature on remote meetings and collaboration is highly dispersed across different fields and outlets. The lack of consensus in the usage of terms and keywords to refer to these phenomena causes numerous papers from one area to slip under the radar of researchers in another.

Given the imperative of building out the post-pandemic hybrid future of work, this conceptual confusion shows a need to find ways to sort and derive principles from the literature. The literature on remote collaboration and meetings is highly dispersed across different fields and outlets, and even within a constrained set of related fields, such as HCI and CSCW, it would be impractical to classify the thousands of papers on hybrid work, such as its effects on productivity, management, collegiality, wellbeing, inclusiveness, etc. As such, in this article we present a systematic literature review (SLR) of the \textit{contexts and tools of hybrid collaboration and meetings} in HCI and CSCW, as represented by those fields' primary database, the ACM Digital Library (ACM DL). 

In this review, \textit{hybrid collaboration} refers to ``collaborative practices that involve simultaneous co-located and remote collaboration with phases of both synchronous and asynchronous work that spans multiple groupware applications and devices''~\cite{neumayr2018} and \textit{hybrid meetings} refers to video- or audio-based communication sessions among co-located and remote participants \cite{saatcci2019hybrid}. We bring them together into the larger whole, \textit{hybrid collaboration and meetings (HCM)}, because they are interrelated (the boundaries grow fuzzier the closer one looks) and because the goal of this article is to unpack how \textit{hybridity} matters when it confers an asymmetry on the coordinated activity.

We used 72 keywords in our search including but not limited to `virtual meetings', `hybrid meetings', `computer-mediated communication' and `video-mediated communication'. After analysis of the 1,209 results retrieved, we selected 62 long and short papers which explore hybrid collaboration and meeting (HCM) settings, regardless of whether the authors explicitly referred to their contexts as `hybrid'. We have classified these papers based on their research questions, methodology, and results, in order to highlight %the boundaries of 
what HCI/CSCW know and do not know about HCM research findings and implementations. The SLR itself covers the ACM until November 2019, to capture the pre-pandemic state of knowledge about HCM, with the extended discussion including relevant 2020 and 2021 publications. That being said, while numerous seminal and valuable publications were discovered in the SLR, we were surprised that the overall composition of the final corpus seemed limited in its capacity to give a complete picture of the phenomena of HCM. As we aim to provide researchers and designers of future HCM systems with a resource to guide their endeavors, we have included discussion of several high-profile publications to complete the ones that were discovered in our systematic approach. Taken together, the goal is to produce a baseline review for post-COVID-19 HCM research and development.

The structure of the article is as follows. \hyperref[sec:partA]{Part A} deals with the outcomes of our systematic approach and comprises the methodology, description, and discussion of results, as well as the suggestion for a taxonomy based on the findings. \hyperref[sec:partB]{Part B} aims at filling in the gaps identified in the systematic review, by providing some high-profile publications that can help us understand missing aspects of hybridity (Sections \ref{sec:missing} and \ref{sec:focal}), and then 
concludes with the key takeaways and proposals for the future of HCM research.

\begin{comment} LONGER STRUCTURE VERSION MAY BE TOO LONG

The structure of the article is as follows: We we start by providing definitions of hybrid collaboration (Section \ref{sec:def_hybrid_collaboration}) and hybrid meetings (Section \ref{sec:def_hybrid_meetings}) and compare the two phenomena (Section \ref{sec:differences}). Then we present the details regarding how we planned and conducted our systematic literature review on HCM in Section \ref{sec:methodology}. These details include our research questions, the queried data sources, our inclusion and exclusion criteria, and information about the search query. We discuss the results of our literature review in terms of thematic overview, terms, citations, venues, authors, domains, study types and data collection in Section \ref{sec:results}. Lastly, we provide a taxonomy of HCM (Section \ref{sec:taxonomy}), listing an overview of the articles concerning group size, number of locations, synchronicity, asynchronicity, software tools and hardware devices. In Section \ref{sec:conclusion}, we conclude with the key takeaways from our systematic literature review and propose future design directions in HCM research. \todo{Revise the structure of the article to match actual contents of the sections in the end OR MAYBE DELETE THIS DUE TO SUGGESTION ABOVE.}

\end{comment}

\subsection{Definition of Hybrid Collaboration}
\label{sec:def_hybrid_collaboration}

\begin{comment} 'PRE-HISTORY' IS A DISTRACTION

%Very early on in their history, humans have discovered the advantages of \textbf{working together} and social cohesion is anchored within the human nature since prehistoric times. Hunters and gatherers worked together to maximize their profits (e.g., by acquiring the meat of large animals they would be unable to hunt down individually) and have a safety net against failure (e.g., during illness, or unsuccessful hunting endeavors where the others of a collective stepped in). In this richness and to this degree, this is special for humans among other living beings because of their unique cognitive capacities and the institution of language that allow for extended communication and coordination of activities. 

%For the scientific perspective adopted throughout this article, the rather imprecise term of `working together' is of course not sufficient but rather the term collaboration can be seen and defined in contrast to other forms of working together such as communication, contribution, coordination, or cooperation. A first step in this direction can be extracted from 

\end{comment}

Denning and Yaholkovsky \cite{denning_2008} hold that collaboration is the ``highest, synergistic form of working together.'' Shah \cite{shah_2010} details how the different forms of working together can be seen in relation to one another suggesting the following clarification to define collaboration and distinguish it from mere cooperation.
\begin{quote}
    ``\textit{Collaboration}. This is a process involving various individuals who may see different aspect of a problem. They engage in a process that goes beyond their own individual expertise and vision to complete a task or project. In contrast to cooperation, collaboration involves creating a solution or a product that is more than the sum of each participant's contribution.''
\end{quote}

For the study of collaborative behavior and interaction between humans, a multitude of models and frameworks have been proposed. One particularly popular example is Johansen's time-space matrix \cite{johansen_1988} that allows for the categorization of groupware or group activities according to their temporal (`same time' vs. `different time') or local (`same place' vs. `different place') nature, resulting in four different quadrants. An example for \textit{same time} (i.e. \textbf{synchronous}), \textit{different place} (i.e. \textbf{{remote}}) would be a telephone call, or a video conference. Other examples would be a meeting in a conference room for \textit{same time, same place} (i.e. \textbf{{co-located}}), an email sent and read at a later point in time for \textit{different time} (i.e. \textbf{{asynchronous}}) and \textit{different place}, or finally an example for \textit{different time} and \textit{same place} would be leaving a note on the kitchen table to be read by some other family member later on. It is important to note that only since the introduction of communication technology is it possible to collaborate at the same time from different places (i.e., \textit{synchronous remote} interaction). 

Modern information and communication technology (ICT) systems enable activities that switch back and forth between all four quadrants of the time-space matrix. For example, a web design team working on a project could be primarily \textit{co-located} in an office while at least one works \textit{remotely} from home. They may use a platform such as Microsoft Teams or Slack to chat  \textit{synchronously} and \textit{asynchronously} or meet \textit{synchronously}, all which may be coordinated with the use of other \textit{asynchronous} resources such as shared files created by the individual members with other tools. Recently, Neumayr et al. \cite{neumayr2018} suggested a definition for hybrid collaboration which incorporates this more modern switching of activities but also delineates what is different from either co-located or remote collaboration: 

\begin{quote}
    ``\textbf{(1.)} Hybrid collaboration switches back and forth between all four quadrants of the time-space matrix. There are constant transitions between co-located and remote as well as synchronous and asynchronous collaboration; \textbf{(2.)} The team size S is greater than just two collaborators and multiple coupling styles can coexist simultaneously within a single team, effectively dividing the whole team in multiple temporary subgroups with each one having a size of $1 \leq S_{sub} \leq S$ and an individual coupling style; \textbf{(3.)} Users typically do not rely on a single groupware application or hardware device but simultaneously use different tools and devices during collaboration.''
\end{quote}

\sr{We will use this definition of hybrid collaboration for the remainder of this article, but given that meetings make up such an important subset of collaboration activity, it is worth also defining hybrid meetings.}
% are a critical aspect of our \textit{hybrid collaboration and meetings} focus.
%treat hybrid collaboration as one part of the subject HCM according to this definition.

\subsection{Definition of Hybrid Meetings}
\label{sec:def_hybrid_meetings}

In the ``meetingization'' of post-industrial societies, meetings rationalize collective social orientation and coordination to work \cite{van_vree_meetings_2011}. From as early as the 1960s, office workers in the U.K. were dedicating half of their time to meeting with others, and only 33 percent of their time was spent alone \cite{stewart1988managers}. Organizational researchers categorize these meetings based on their topics, formats, and goals, ranging from decision-making to brainstorming \cite{allen2014understanding}. The general condition of meeting is ``a gathering of two or more people for purposes of interaction and focused communication'' \cite{volkema1995organizational, goffman1961encounters, schwartzman1989meeting}. As teleconferencing technologies gave rise to the concept of `tele-commuting' in the 1970s, researchers began to explore the notion of \textit{gathering} from functional and technological standpoints. They explored which functions relied most and least on all participants gathering in person \cite{pye1976effect, pye1978}, and which technologies (alone or in combination) facilitated remote and hybrid gatherings -- with specific emphasis on whether video would be valuable in remote meetings or audio alone would be adequate \cite{ochsman_effects_1974, pye&williams1977}. Since then, research into telepresence for meetings has explored many variations in both social and techological factors \cite{finn_video-mediated_1997, harrison2009, olson_working_2013, kraut_intellectual_2014, lombard2015immersed}.

The earliest term closest to hybrid meetings in academic literature is `partially distributed meetings'. However, that term does not fully imply whether the distribution is merely physical or temporal, and whether `partially' refers to some sections of meeting being fully co-located or remote rather than the condition that co-located and remote participants are attending at the same time. For the purposes of this review, we define hybrid meetings as video- or audio-based meetings among co-located and remote participants~\cite{saatcci2019hybrid}. This definition is agnostic to the distribution of participants in time or space. 

\subsection{Differences and Similarities between Hybrid Collaboration and Hybrid Meetings}
\label{sec:differences}

One problem of defining hybridity for the purposes of an SLR is that the concept of hybrid meetings can be regarded as both more general \textit{and} more particular in comparison to hybrid collaboration. 

It is more general insofar as not all aspects of hybrid collaboration's definition (see Section~\ref{sec:def_hybrid_collaboration}) must apply for a meeting to be regarded as a hybrid meeting. For example, concerning part \textbf{(1.)} of that definition, it is entirely valid to only rely on synchronous communication during a hybrid meeting, concerning part \textbf{(2.)} there are often fewer tendencies to form several subgroups in meetings (due to the synchronized nature of communication), and concerning part \textbf{(3.)} it is oftentimes the case that only one or very few tools and devices are used as opposed to the typical practices in hybrid collaboration where usually multiple tools and also personal devices are an integral part of the collaboration.

On the other hand, it is more particular, insofar as the typical collaborative activities found in hybrid meetings usually concern---as the name states---meeting activities and therefore a subset of activities and tasks as compared to the broader term of collaboration. For example, hybrid collaboration can include all of co-authoring, co-programming, co-design, sense-making and many more in a hybrid team context whereas hybrid meetings typically have activities more similar to traditional meetings, such as information seeking, generation/discussion of ideas, delegation of work, or presentations, just to name a few.

There are also differences in the temporal scope of hybrid meetings and hybrid collaboration. A hybrid meeting is mostly limited to one occurrence of people coming and working together for usually a few hours at maximum. Conceptually, when talking about recurring (e.g., weekly) meetings, we see this rather as repetitions of single instances of meetings instead of a single longer meeting process. A hybrid collaboration, however, could last for days, weeks, months or even years in theory (e.g., a paper co-authoring process can often take several months).

In conclusion, though, while typically differing in the tasks, synchronicity, tendency to form subgroups, and permanence \cite{lee15}, hybrid collaboration and hybrid meetings share a common notion of %the entire part of the definition that defines
collective work by teams comprised of both co-located local and remote participants. As such, we believe that a joint literature review will inform us about what has been already studied and what needs to be further researched. 

\subsection{Aim: Understanding Hybridity}

\sr{We said above that we are exploring hybrid collaboration and meetings (HCM) together, because our goal is to unpack how \textit{hybridity} matters when it confers an asymmetry on the coordination that occurs within the interrelated concepts of collaboration and meetings. 'Hybrid' is the key concept under investigation because it has become the term \textit{du jour} for several co-located participants engaging in coordinated work with one or more remote participants. While it has its own looseness, it is perhaps at least more connotative of the asymmetrical cohort distribution than other terms that have been used, such as `online', `distributed' or `partially distributed'. Those terms may be confused with either fully virtual collaboration (e.g.~\cite{olson_working_2013}) or even in-person collaboration (e.g.~\cite{marquardt_cross-device_2012}). In some ways such terms are products of a time before `hybrid' itself came to prominence in this context. That being said, `hybrid' is an established term in research on coordinated work contexts, especially in education, where `hybrid education' \cite{hall2015hybrid} and `hybrid learning' \cite{chen2014learning} were seen as solutions to problems of scale and inclusion before the COVID-19 pandemic.} 
%, and hybrid settings may become more common in education if social distancing regulations continue. Since hybridity in education refers 
%, to refer to the merging of in-person and online education.
%and there is a shared understanding around this definition, calling the meetings and collaboration of co-located and remote participants as `hybrid' is associative for scholars from different disciplines. 
%Further, the communication technology and event industries also refer to this phenomenon as `hybrid'\footnote{What are hybrid meetings? Why should you prepare your offices for them? \url{https://www.barco.com/en/news/2020-11-03-what-are-hybrid-meetings}, last access February 15, 2021.}. 

\sr{Our primary goal is to help future researchers, designers, developers, and others \textit{understand how hybridity matters} to the tools and processes of collaboration and meetings. Our secondary goal is to help those \textit{searching for prior literature} on hybrid collaboration and meetings to understand what they will and will not find when undertaking such a search. This is important because, as we have noted above and will discuss further, there is a large difference between what can be found using the term and what is relevant to the concept. So, for example, the SLR revealed some work on document collaboration when it specifically investigated how it matters in context of hybridity, but this does not represent total coverage of the document collaboration literature. Another example, perhaps controversial, is that the term `Media Space' was not included in the SLR, partially because it would result in many false positives\footnote{The words `media' and `space' are often used together in metaphorical reference to both mass media and social media, as well as appearing adjacent but without combined meaning in the ACM DL. These vastly increase the number of false positives. We could have limited the use of the term to the title only, but that would deviate from the SLR methodology and, of course, not all research on media spaces include the term in their title.}, partially because some contributions to Media Space research do not appear in the ACM DL, and partially because of the nature of Media Space research itself (as we will discuss later). To remediate this exclusion, we have included Media Space research in our discussion on what is missing (Section \ref{sec:missing}). Given these goals, one obvious outcome of this article will be, as those before us have urged~\cite{olsonolson_vocabulary_1997, schmidt_problem_2002} to emphasize the need for a unifying vocabulary to make future research more easily discoverable, aggregated, and comparable.}

%\clearpage
\vspace{1.5cm}
{\Large \textit{\textbf{PART A - Systematic Findings}}}
\label{sec:partA}

\section{Methodology}
\label{sec:methodology}
The following sections present an overview of our systematic literature review's process. We used an approach that was initially inclusive regarding the search terms to obtain a broad view of the potentially relevant literature and not miss related publications. To achieve this, we designed our search query to be comparatively permissive by using a broad set of search terms, although often in systematic literature reviews, researchers only query for one or very few concrete terms. We knew after pretesting our query, that we might have to deal with a large number of semantically false positives by doing so, but found it necessary due to the conceptual confusion of the terms used to describe the phenomena of HCM. In a second step, we manually applied rather strict criteria for selecting the publications into the final corpus of relevant literature. 

\subsection{Planning the Review}
In this section, we describe how we planned our systematic literature review by presenting our research questions, justifying our queried data sources and discussing the inclusion and exclusion criteria we applied.

\subsubsection{Research Questions}
In our previous work, we noticed a lack of a systematic overview of the related literature on HCM. Our general research goal, to provide such a systematic overview of existing work on HCM, is motivated by a desire to provide future researchers with the most relevant dimensions and characteristics of HCM research in the ACM DL. \tn{SLRs often have a rather narrow focus on one specific delimitable characteristic or dimension, for example on Social Presence in virtual environments \cite{oh2018} (here the authors additionally focus on ``predictors of social presence''), on technologies which support awareness in collaborative systems \cite{lopez2017}, or on explanations in decision support and recommender systems \cite{nunes2017}. This focused approach allows these SLRs to distill important predictors for the dimensions or measurements for success. With our aim of giving a broader overview of the concepts of HCM, we expected that the obtained references will contain many different subsets of thematic foci on dimensions (e.g., on predictors, on awareness, etc.) all with a connection to HCM in their core. This, in turn, implies a different scope of our SLR, where not individual dimensions and characteristics of HCM are in the center, but the concept of HCM in its entirety, containing a rich diversity of different dimensions and characteristics that were investigated in this context. Therefore, we focused more on the actual settings that were examined and what could be learned from them as well as the historical evolvement of HCM.} \tn{The lowest common denominator among the systematically obtained publications is the fact that they all involve a mixture of remote and co-located individuals who collaborate or meet in a hybrid setting while many different research questions are covered. A uniform possibility to describe all HCM involving studies in hybrid settings (representing such a common denominator) is suggested in the form of a taxonomy in Section \ref{sec:taxonomy}. Furthermore, to cover a meta-level of information about previous work on HCM and its historical evolvement, and inspired by \cite{nunes2017} and \cite{neumayr2020}, who broke down their overall aim into several more specific research questions, we conceived of the following concrete sub-questions to be answered by the systematic literature review}:
\begin{itemize}
    \item Research Question 1 (RQ1): Is there research which deals with HCM according to our definitions (see Sections \ref{sec:def_hybrid_collaboration} and \ref{sec:def_hybrid_meetings}) and if yes, where is the work on HCM \textit{thematically rooted}?
    \item RQ2: How were HCM \textit{called/coined/termed}?
    \item RQ3: Which \textit{research topics} have been covered and which \textit{research methods} have been used when studying HCM?
    %\item RQ6: How do \textit{the results differ} and (how) can they \textit{be categorized}?
    \item RQ4: Was HCM discussed using the term \textit{`hybrid'} or at least with the aim of researching settings that can be regarded as HCM \textit{(implicit/explicit)}?
    \item RQ5: What \textit{domains} are \textit{relevant} for HCM and what \textit{domains make use} of HCM?
    \item RQ6: Since when is research on HCM reported and how did it \textit{chronologically evolve}; are there \textit{historical shifts} or \textit{trends} in terms of \textit{`hybridity'} in HCM?
    \item RQ7: How can work on HCM beyond the mere results be \textit{thematically clustered} and \textit{categorized}?
\end{itemize}

\subsubsection{Queried Data Sources}
The ACM DL\footnote{\url{https://dl.acm.org/about}, accessed January 21st, 2021} is one of the most extensive databases (approx. 2.85 million publications) for scientific literature in the computing domain with a temporal coverage of the publication years starting from 1908. On the one hand, we decided to use the ACM DL mainly because of its rather broad spectrum of computing and technology-related research topics giving us the possibility to cover many different disciplines in which HCM potentially play a role (such as \textit{Human-Centered Computing} with subfields HCI and CSCW, \textit{Security and Privacy}, or more technically-centered fields like \textit{Networks}, \textit{Information Systems} or \textit{Software and its Engineering} to just name a few of ACM DL's potentially relevant CCS categories\footnote{\url{https://dl.acm.org/ccs}, accessed January 21st, 2021}). On the other hand, we decided to not use an even broader database, such as Google Scholar\footnote{\url{https://scholar.google.com}, accessed January 21st, 2021}. This is because first experiments with our query on Google Scholar resulted in several tens of thousands of search results of which we examined a random sample leading us to the expectation to retrieve a very high percentage and, therefore, an unmanageable amount of non-relevant publications with the additional danger of retrieving such publications that are of inferior quality or not published under peer-review procedures.

Although we are aware that by this decision no claim for completeness is possible, our aim is, nevertheless, to give a well-balanced overview of relevant literature on HCM. \textbf{Concerning the field of HCM}, we regard the ACM DL as a very broad database from a disciplinary point of view as compared to other databases, such as IEEE Xplore\footnote{\url{https://ieeexplore.ieee.org}, accessed January 21st, 2021}, at least judging from samples taken from pretests done with our query. In other words, according to the pretesting of our query, the ACM DL is characterized by a more horizontal quality of the results for HCM (covering more different disciplines where HCM might play a role), while IEEE Xplore was rather more vertically oriented for our sample of HCM-related keywords (covering fewer different disciplines with a higher number of publications in each).

All of these considerations in addition to our awareness of other literature reviews in similar fields of research either using the ACM DL as one of few major data sources (see e.g., \cite{nunes2017}), exclusively utilizing it (see e.g., \cite{brudy2019, neumayr2020}) or focusing on specific conference proceedings (such as CHI, CSCW, or CSCL, see, e.g., \cite{keegan2013,liu2014,tang2014,wallace2017}) led to the decision to rely on the ACM DL exclusively.

\subsubsection{Inclusion and Exclusion Criteria}
The basic premise of our systematic literature review was to include the largest possible share of research in the ACM DL dealing with HCM according to any of our definitions (see Sections \ref{sec:def_hybrid_collaboration} and \ref{sec:def_hybrid_meetings}), even if the authors did not use the according terms to describe their work.

Our inclusion criteria as reflected by our search query (see Section \ref{sec:search_query}) can, therefore, be defined as follows:

\begin{itemize}
    \item IC1: The publication contains research about \textbf{hybrid collaboration} as defined in Section \ref{sec:def_hybrid_collaboration}.
    \item IC2: The publication contains research about \textbf{hybrid meetings} as defined in Section \ref{sec:def_hybrid_meetings}.
\end{itemize}

We included papers if at least one of the two inclusion criteria applied.
Regarding our exclusion criteria (see below), we excluded papers if at least one of them applied.

\begin{itemize}
    \item EC1: The publication is not relevant because it does not deal with HCM, therefore, constituting a false positive.
    \item EC2: The publication is not a research paper (or journal article), e.g., it is an abstract for a panel, a workshop invitation or the like. Please note, that we also included short papers, extended abstracts or late-breaking work to not miss potentially relevant research that might have been abandoned since then.
    \item EC3: The publication language is not English. We initially discussed also considering papers in Turkish or German because these are the \tn{two} main authors' first languages but concluded that a joint and equitable interpretation of the papers would be difficult if only one of the main authors could understand the contents of the respective papers.
\end{itemize}

\tn{To further illustrate and make our procedure transparent, \autoref{tab:exclusions_2019} gives an overview of a subset of publications and justifies why they were excluded based on the EC presented above. This table includes all originally selected publications of the year 2019. To our surprise, we had to exclude all of the 2019 publications, despite our systematic approach's search terms being quite general (e.g. as \textit{computer-mediated communication}, or \textit{virtual collaboration} (see Section \ref{sec:search_query})). \autoref{tab:exclusions_2019} shows the title, reference and publication title of all such exclusions and shows one possible string which had triggered the search query. The table presents details about why the publications were excluded. Most of the publications rather obviously did not fulfill our criteria, which was often evident just by reading the abstract and the title. For example, our search terms, such as ``hybrid event'' \cite{almeida_hierarchical_2019} or ``virtual communcation'' \cite{augustine_distributed_2019} were understood as technical terms among software or hardware components and not about human communication or collaboration behavior in some of the publications (e.g., in \cite{almeida_hierarchical_2019, augustine_distributed_2019, wang_gpu-based_2019, wang_parallelizing_2019}). Some further publications focused on dyads in their study which \textit{a priori} prevents the mixture of co-located and remote participants inherent to HCM (e.g., \cite{ahmed_structuring_2019, griggio_augmenting_2019, guo_as_2019}). However, there were also some more plausible candidates where the decision for exclusion or non-exclusion was not clear only by reading title and abstract. For example, in \cite{alvarez_virtual_2019}, judging from the descriptions in the full text, each of 25 participating students were part of a different global virtual team, therefore, no co-located activities were covered and none are reported. Another example is \cite{duan_increasing_2019} where it became only apparent after reading the descriptions about the study setup that the participants were in ``separate soundproof rooms'' \cite[p.8]{duan_increasing_2019} that no co-located portion was investigated. In a similar fashion, we analyzed and discussed the contents of each publication which seemed to be a plausible candidate based on the title and abstract on our mission to find the HCM needles in the haystack.}

\renewcommand{\arraystretch}{1.7}
\begin{longtable}[c]{p{3cm}p{0.5cm}p{2.5cm}p{2cm}p{4cm}}
    % \centering
    % \begin{tabular}
    \caption{Details for the exclusion of all selected papers of the year 2019.\label{tab:exclusions_2019}}\\
    \toprule
        \textbf{Title} & \textbf{Ref} & \textbf{Publication \mbox{Title}} & \textbf{Possible Query \mbox{Trigger}} & \textbf{Details for Exclusion} \\ \midrule
        \endfirsthead
        \textbf{Title} & \textbf{Ref} & \textbf{Publication \mbox{Title}} & \textbf{Possible Query \mbox{Trigger}} & \textbf{Details for Exclusion} \\ \midrule
        \endhead
        \multicolumn{5}{ r }{\textit{(Continued)}}\\
        \endfoot
        \endlastfoot
        Structuring Online Dyads: Explanations Improve Creativity, Chats Lead to Convergence & \cite{ahmed_structuring_2019} & Proceedings of the 2019 on Creativity and Cognition & Author tag: "computer-mediated", "communication" & Study with dyads in remote collaboration and not hybrid. \\ 
        A Hierarchical Architectural Model for Network Security Exploring Situational Awareness & \cite{almeida_hierarchical_2019} & Proceedings of the 34th ACM/SIGAPP Symposium on Applied Computing & Abstract: "hybrid", "event" & Technical paper, situational awareness regards not humans but computer components, "hybrid event" is understood as a software message. \\ 
        The virtual collaborative work and the development of intercultural competences in university student's: The case of Virtual Global Teams & \cite{alvarez_virtual_2019} & Proceedings of the Seventh International Conference on Technological Ecosystems for Enhancing Multiculturality  - TEEM'19 & Title: "Virtual", "Teams" & Global virtual teams (GVT) were investigated, judging from the descriptions they were remote only (each of 25 students in another GVT). \\ 
        Distributed Computation in Node-Capacitated Networks & \cite{augustine_distributed_2019} & The 31st ACM on Symposium on Parallelism in Algorithms and Architectures & Abstract: "virtual", "communication" & Technical paper about computer networking. Communication is understood as among network nodes and not humans. \\ 
        Understanding Digitally-Mediated Empathy: An Exploration of Visual, Narrative, and Biosensory Informational Cues & \cite{curran_understanding_2019} & Proceedings of the 2019 CHI Conference on Human Factors in Computing Systems & Author tag: "computer-mediated", "communication" & Not a collaboration example, experiment tries to find out which measures increase empathic accuracy of observers of a VR video (e.g., only seeing the video as baseline versus subtitle of how the person feels versus electrodermal acticity). \\ 
        Geollery: A Mixed Reality Social Media Platform & \cite{du_geollery:_2019} & Proceedings of the 2019 CHI Conference on Human Factors in Computing Systems & Abstract: "virtual", "meetings" & This is a purely remote approach to letting ``remote participants'' [Abstract, p. 1] chat in virtual environments that are spatially aware (such as Google Street View). \\ 
        Increasing Native Speakers' Awareness of the Need to Slow Down in Multilingual Conversations Using a Real-Time Speech Speedometer & \cite{duan_increasing_2019} & Proceedings of the ACM on Human-Computer Interaction & Author tag: "computer-mediated", "communication" & Remote only, participants in "separate soundproof rooms" [p. 8]. \\ 
        Is Technology Killing Human Emotion?: How Computer-Mediated Communication Compares to Face-to-Face Interactions & \cite{eddy_is_2019} & Proceedings of Mensch und Computer 2019 on   - MuC'19 & Title: "Computer-Mediated", "Communication" & Online survey, isolated investigation of behavior during either face-to-face or remote communication. \\ 
        Facial Cues for Deception Detection in Virtual Reality Based Communication & \cite{farizi_facial_2019} & Proceedings of the 3rd International Conference on Big Data and Internet of Things  - BDIOT 2019 & Title: "Virtual", "Communication" & Study setting is not collaborative, but a lab experiment where single participants are presented with either a 2D video or a 3D avatar translated from the video and judge their deception behavior [p. 66f]. \\ 
        Managerial Visions: Stories of Upgrading and Maintaining the Public Restroom with IoT & \cite{fox_managerial_2019} & Proceedings of the 2019 CHI Conference on Human Factors in Computing Systems & Abstract: "computer-mediated", "collaboration" & Not about collaboration as we understand it, but rather about managers installing and workers using IoT technology in public restrooms. \\ 
        Author Highlights for the Past 35 Years: An Analysis of the Most-Published Authors and Most-Cited Papers in The DATA BASE for Advances in Information Systems & \cite{gallivan_author_2019} & SIGMIS Database & Abstract: "virtual", "teams" & Anniversary paper presenting the most cited papers of a specific journal, no original research. \\ 
        Customizations and Expression Breakdowns in Ecosystems of Communication Apps & \cite{griggio_customizations_2019} & Proceedings of the ACM on Human-Computer Interaction & Author tag: "computer-mediated", "communication" & Interview study with 15 ``extreme users'' of messaging apps, such as WhatsApp or Telegram (therefore, remote only). \\ 
        Augmenting Couples' Communication with Lifelines: Shared Timelines of Mixed Contextual Information & \cite{griggio_augmenting_2019} & Proceedings of the 2019 CHI Conference on Human Factors in Computing Systems & Author tag: "computer-mediated", "communication" & Study with couples (therefore, dyads), not hybrid. \\ 
        As If I Am There: A New Video Chat Interface Design for Richer Contextual Awareness & \cite{guo_as_2019} & Extended Abstracts of the 2019 CHI Conference on Human Factors in Computing Systems & Author tag: "computer-mediated", "communication" & Dyads in purely remote video chat, not hybrid. \\ 
        On the Internet, Nobody Knows You'Re a Dog... Unless You'Re Another Dog & \cite{hirskyj-douglas_internet_2019} & Proceedings of the 2019 CHI Conference on Human Factors in Computing Systems & Abstract: "computer mediated", "communication" & Dog interfaces for animal to animal communication, not hybrid. \\ 
        Is Seeing Believing?: The Effect of Morphological Congruent Visual Feedback on Mediated Touch Experience & \cite{ipakchian_askari_is_2019} & Extended Abstracts of the 2019 CHI Conference on Human Factors in Computing Systems & Author tag: "computer mediated", "communication" & Remote only, ``geographically separated individuals'' are in the focus [Abstract]. \\ 
        AI-Mediated Communication: How the Perception That Profile Text Was Written by AI Affects Trustworthiness & \cite{jakesch_ai-mediated_2019} & Proceedings of the 2019 CHI Conference on Human Factors in Computing Systems & Abstract: "Computer-Mediated", "Communication" & Communication and not collaboration is studied, only online (remote). \\ 
        Technological Frames and User Innovation: Exploring Technological Change in Community Moderation Teams & \cite{kiene_technological_2019} & Proceedings of the ACM on Human-Computer Interaction & Author tag: "computer-mediated", "communication" & Online communities (such as Reddit and Discord) are studied, no particular instance of collaboration is investigated. \\ 
        Developing a Hand Gesture Recognition System for Mapping Symbolic Hand Gestures to Analogous Emojis in Computer-Mediated Communication & \cite{koh_developing_2019} & ACM Trans. Interact. Intell. Syst. & Abstract: "computer-mediated", "communication" & The messaging part of CMC is in the focus, remote only. \\ 
        Does Social Sensitivity Impact Virtual Teams? & \cite{lacher_does_2019} & Proceedings of the 50th ACM Technical Symposium on Computer Science Education & Title: "Virtual", "Teams" & Focus on purely virtual teams. Judging from the descriptions, teams only collaborated via Discord and did not work co-locatedly. While there may be phases of F2F collaboration in between, this was not reported. \\ 
        Tom-Talker: Pet Robot Social Incentive System for Urban Autism & \cite{lou_tom-talker:_2019} & Proceedings of the 2019 International Electronics Communication Conference on   - IECC '19 & Abstract: "virtual", "communication" & Co-located or person-to-robot interaction only. \\ 
        Pet Robot Emotional Interaction for Urban Autism & \cite{lou_pet_2019} & Proceedings of the 2019 2nd International Conference on Intelligent Science and Technology  - ICIST 2019 & Abstract: "virtual", "communication" & Co-located or person-to-robot interaction only. \\ 
        Towards Collaborative Photorealistic VR Meeting Rooms & \cite{schafer_towards_2019} & Proceedings of Mensch und Computer 2019 on   - MuC'19 & Abstract: "virtual", "meeting" & VR, remote only, no study yet. \\ 
        Democratic power structures in virtual communities & \cite{seidel_democratic_2019} & Proceedings of the 24th European Conference on Pattern Languages of Programs  - EuroPLop '19 & Abstract: "virtual", Index Terms: "Collaborative", Abstract: "systems" & Suggestion of design patterns for virtual communities, remote only. \\
        Role of Technology in Multicultural Environment: Impact of MOODLE Learning System on Global Virtual Team Performance & \cite{shahid_role_2019} & Proceedings of the 2nd International Conference on Big Data Technologies  - ICBDT2019 & Title: "Virtual", "Team" & Investigation of global virtual teams and usage of moodle, remote only. \\ 
        Accessible Video Calling: Enabling Nonvisual Perception of Visual Conversation Cues & \cite{shi_accessible_2019} & Proceedings of the ACM on Human-Computer Interaction & Author tag: "computer-mediated", "communication" & One-on-one conversations between sighted and limited sight or no vision persons, remote only. \\ 
        Do people virtually support their favorite cricket team?: insights from 2018 Asia cup & \cite{singh_people_2019} & Proceedings of the Third International Conference on Advanced Informatics for Computing Research  - ICAICR '19 & Abstract: "virtual", "team" [referring to cricket team] & Social Media analysis of (Twitter) tweets, remote only. \\ 
        Trusted Teammates: Commercial Digital Games Can Be Effective Trust-Building Tools & \cite{tan_trusted_2019} & Extended Abstracts of the Annual Symposium on Computer-Human Interaction in Play Companion Extended Abstracts  - CHI PLAY '19 Extended Abstracts & Abstract: "virtual", "teams" & Remote only, experiment gathered virtual teams over Google Hangouts. \\ 
        How to Communicate when Submitting Patches: An Empirical Study of the Linux Kernel & \cite{tan_how_2019} & Proceedings of the ACM on Human-Computer Interaction & Abstract: "computer-mediated", "communication" & Remote only, analysis of online documents and emails. \\ 
        Parallelizing cryo-EM 3D reconstruction on GPU cluster with a partitioned and streamed model & \cite{wang_parallelizing_2019} & Proceedings of the ACM International Conference on Supercomputing  - ICS '19 & Abstract: "hybrid", "communication" & Technical paper, where (hybrid) communication is only understood as communication between computing memory \\ 
        GPU-based 3D cryo-EM Reconstruction with Key-value Streams: Poster & \cite{wang_gpu-based_2019} & Proceedings of the 24th Symposium on Principles and Practice of Parallel Programming & Abstract: "hybrid", "communication" & Technical paper, where (hybrid) communication is only understood as communication between computing memory \\ 
        Culturally-Embedded Visual Literacy: A Study of Impression Management via Emoticon, Emoji, Sticker, and Meme on Social Media in China & \cite{wang_culturally-embedded_2019} & Proceedings of the ACM on Human-Computer Interaction & Abstract: "computer-mediated", "communication" & Interview study with 30 social media users in China, remote only. \\ 
        Gender Effects on Collaborative Online Brainstorming Teamwork & \cite{yuan_gender_2019} & Extended Abstracts of the 2019 CHI Conference on Human Factors in Computing Systems & Abstract: "computer-mediated", "communication" & 2 experiment conditions to find out about gender differences in group compositions: one is face-to-face, the other is purely online. \\ 
        Managing Stress: The Needs of Autistic Adults in Video Calling & \cite{zolyomi_managing_2019} & Proceedings of the ACM on Human-Computer Interaction & Abstract: "computer-mediated", "communications" & Interview study with ``autistic adults about their perceptions of'' video conferencing compared to other CMC or face-to-face, no particular collaboration scenario. \\ \bottomrule
\end{longtable}
\renewcommand{\arraystretch}{1}

% \begin{figure}
%   \includegraphics[width=\linewidth]{figures/exclusions_2019.pdf}
%   \caption{Details for the exclusion of all selected papers of the year 2019.}}
%   \label{fig:exclusions_2019}
% \end{figure}

\subsection{Conducting the Review}
In this section, we describe how we conducted the review by detailing on the search terms that were used and the results we obtained through the query (RQ1).

\subsubsection{Search Query}
\label{sec:search_query}
We queried the ACM DL on November 8th, 2019 and searched the ACM Full-Text Collection. In order to cover all relevant research, even if it was not named exactly `hybrid collaboration' or `hybrid meeting', we used a number of more general terms in combination. The keywords used \tn{(also see \autoref{tab:search_query})} consisted of two sets, \tn{set Adjectives and set Nouns,} that were combined with a logical AND operator, while items within the two sets \tn{(i.e., within Adjectives and within Nouns)} were connected through logical OR operators. The set \tn{Adjectives} consisted of the keywords: \textit{hybrid, partially distributed, virtual, video-based, video-mediated,} and \textit{computer-mediated}. The set \tn{Nouns} consisted of the keywords: \textit{collaboration, event, meeting, team, communication,} and \textit{collaborative system} \tn{as well as their plural forms}. We were aware that this search query potentially would return a high percentage of false positives (e.g., a paper returned because it contains `virtual' and `communication' but not dealing with HCM \textit{per se}) but our intention was to cover as much relevant research as possible and not losing work due to keyword mismatches. To account for different spellings or word stems, the ACM DL implicitly provided stemming support. Please note that ``[t]he new Digital Library is using a different search engine'' to the one used in November 2019 according to information received by the ACM upon a related request. 

\begin{table}
    \centering
    \caption{Keywords contained in our search query. Terms within Adjectives and Nouns were connected by logical OR operators. The two sets were then joined with a logical AND operator, leading to 72 different combinations (including plural forms). The search query triggered if at least one element from each set was found.}
    \begin{tabular}{ll}
    \toprule
        Adjectives & Nouns \\ \midrule
        hybrid & collaboration(s) \\ 
        partially distributed & event(s) \\ 
        virtual & meeting(s) \\ 
        video-based & team(s) \\
        video-mediated & communication(s) \\
        computer-mediated & collaborative system(s) \\ \bottomrule
    \end{tabular}
    \label{tab:search_query}
\end{table}

Additionally, we applied two refinements that can be regarded as a filtering mechanism for the results returned. The first one was called ``Published by: ACM'' (now called ``Publisher: Association for Computing Machinery'') in order to filter out potentially marginal publications from adjunct publishers whose quality is difficult to evaluate \textit{a priori}. The second one defined the ``Content Formats: PDF'' (this is now called ``Media Formats'' in the new ACM DL) to ensure that the resulting publications can be interpreted uniformly.

\subsubsection{Query Results}
\begin{figure}
  \includegraphics[width=.7\linewidth]{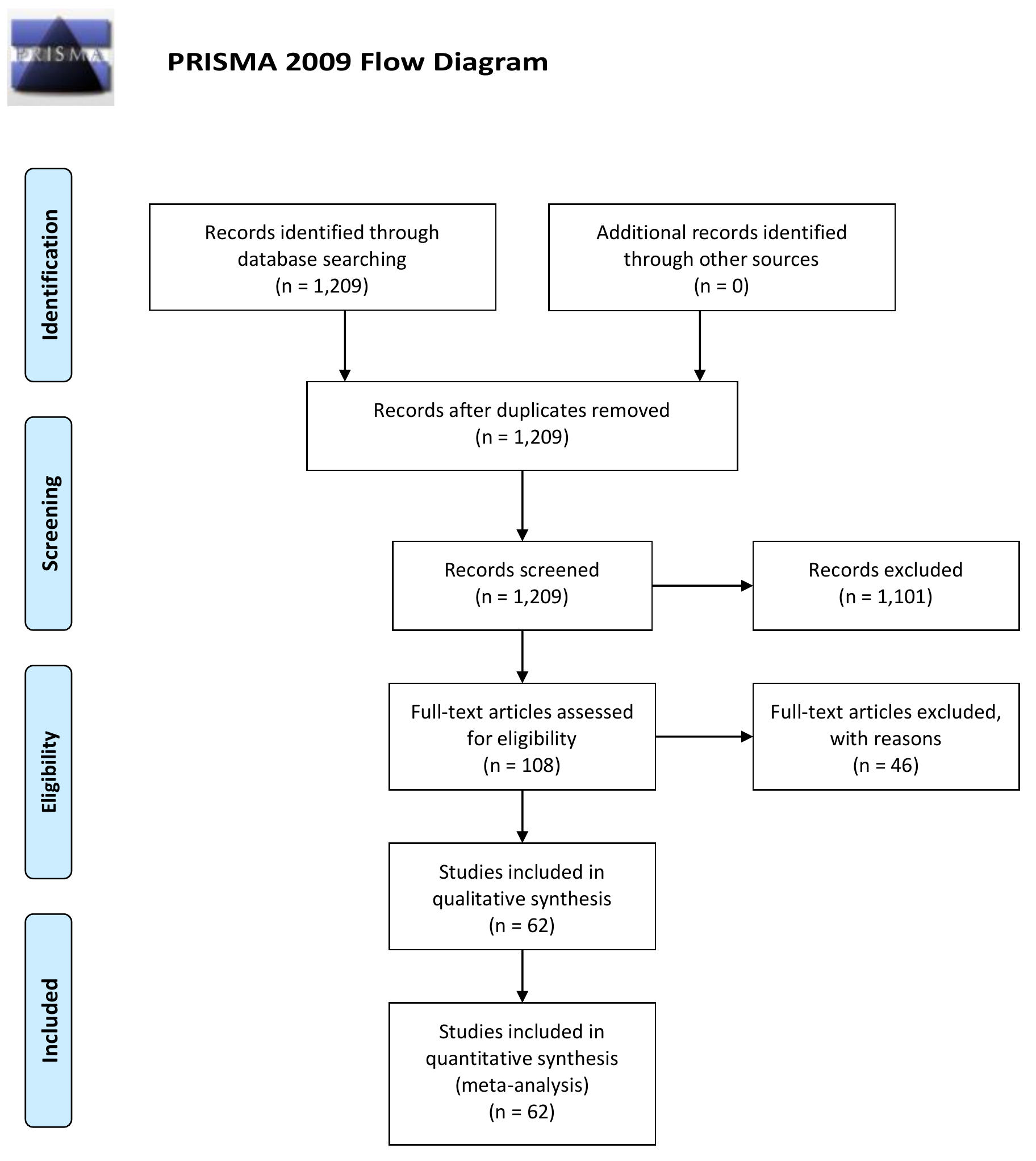}
  \caption{PRISMA flow diagram according to \cite{moher09}.}
  \label{fig:prisma}
\end{figure}

Overall, we obtained 1,209 results spanning the years 1982 to 2019 (up until November). The publications of the years 2018 and 2019 (71 papers) were judged and thoroughly discussed together by the two main authors to adjust our judgements. Next, the result set was split into odd and even years and one main author judged the even and the other the odd years' publications.
We ended up with 44 publications regarded as fulfilling the inclusion criteria and 64 publications for further discussion leading to 108 potentially relevant publications (8.93\% of the overall result set). After further discussions between the authors, 18 of the 64 publications that were marked for further discussions were regarded as fulfilling the inclusion criteria and \textit{not} fulfilling the exclusion criteria. This resulted in a final set of 62 relevant publications or 5.13\% of the overall result set (approximately 95\% false positives) which \tn{is a lower but not entirely unsimilar rate than comparable, initially inclusive open-database approaches achieved, e.g., 90\% reported in \cite{neumayr2020} or 82\% reported in \cite{nunes2017}. The high rate of false positives in our SLR can be explained by the terminological ambiguity with which prior research has dealt with the topics of HCM, which made it necessary to initially use a broad query and iteratively sort out unrelated papers, as we will further discuss throughout the article. \autoref{tab:exclusions_2019} provides a lively illustration of why we had to exclude such a high number of papers.}
Concerning our RQ1, many of the publications have an HCI and/or CSCW focus which shows that research on HCM is rooted in these two fields. For an overview of our inclusion and exclusion process, see the PRISMA flow diagram in Figure \ref{fig:prisma}.

\section{Results}
\label{sec:results}
The earliest paper in our final result set is a 1988 paper \cite{ahuja_rapport_1988} about a multimedia conferencing system called ``Rapport'' which already used networked voice communication and provided a shared workspace on interconnected Sun workstations. This is the only paper from the 1980s, while seven publications are from the 1990s. 20 papers are from the 2000s and 34 papers are from the 2010s (up until such published and indexed by the ACM  before November 8th, 2019). The year with the most publications in our final set is 2012 with eight publications, followed by 2011 (six) and 2013 (five). All other years resulted in four or fewer publications in our final set. Interestingly, all 34 originally selected publications of the year 2019 (including 9 journal articles) did not make it into our final set due to our inclusion or exclusion criteria (also see \autoref{tab:exclusions_2019}). Similarly, 14 papers of the 1980s were excluded. Of the 108 publications of the 1990s in our original set only seven were regarded as relevant. These observations can be seen as first answers to our \tn{RQ6}, which will be further detailed in the taxonomy presented in Section \ref{sec:taxonomy}.

\subsection{Thematic Overview}

\begin{figure}
  \includegraphics[width=\linewidth]{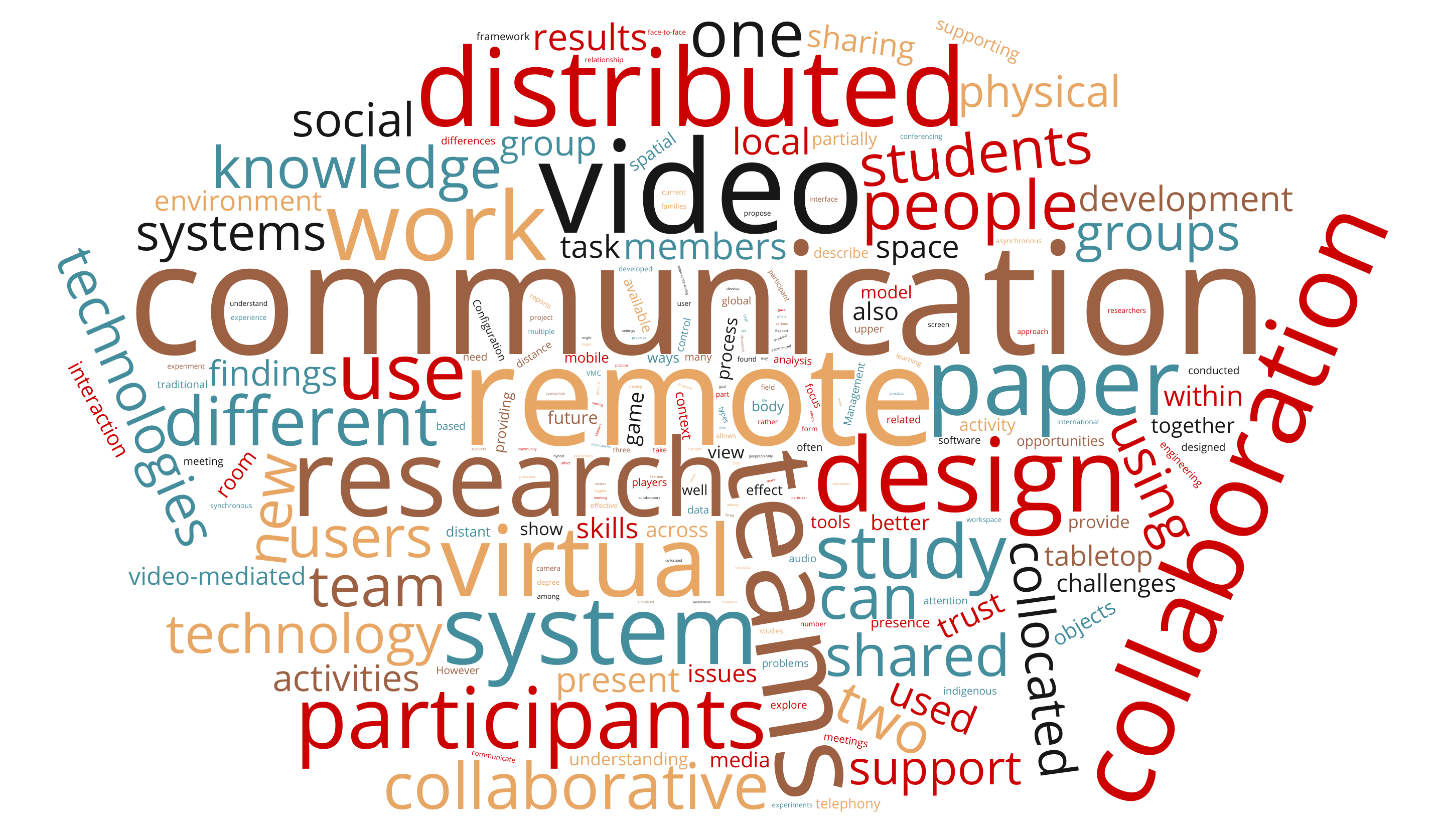}
  \caption{Wordcloud with all the words as found in the abstracts accounting for such words that were found at least five times. Typical stop words were removed. Credit to wortwolken.com.}
  \label{fig:wordcloud}
\end{figure}

\begin{table}[]
\begin{tabular}{llllll}
\multicolumn{1}{l}{\bfseries Word} & \multicolumn{1}{l}{\bfseries Frequency} & \multicolumn{1}{l}{\bfseries Word} & \multicolumn{1}{l}{\bfseries Frequency} &
\multicolumn{1}{l}{\bfseries Word} &
\multicolumn{1}{l}{\bfseries Frequency} \\ \hline
communication & 56        & shared       & 21        & results        & 14        \\ 
remote        & 45        & students     & 21        & sharing        & 13        \\ 
video         & 42        & team         & 21        & space          & 13        \\ 
teams         & 41        & two          & 21        & task           & 13        \\ 
distributed   & 36        & users        & 21        & trust          & 13        \\ 
research      & 36        & technologies & 20        & environment    & 12        \\ 
collaboration & 34        & technology   & 20        & skills         & 12        \\ 
design        & 33        & using        & 20        & tabletop       & 12        \\ 
virtual       & 33        & collocated   & 19        & within         & 12        \\ 
system        & 32        & groups       & 19        & also           & 11        \\ 
work          & 32        & systems      & 19        & challenges     & 11        \\ 
paper         & 31        & social       & 18        & issues         & 11        \\ 
participants  & 29        & support      & 18        & video-mediated & 11        \\ 
use           & 27        & physical     & 17        & across         & 10        \\ 
study         & 26        & used         & 16        & better         & 10        \\ 
different     & 24        & activities   & 15        & game           & 10        \\ 
people        & 24        & local        & 15        & interaction    & 10        \\ 
collaborative & 23        & members      & 15        & media          & 10        \\ 
one           & 23        & development  & 14        & objects        & 10        \\ 
can           & 22        & findings     & 14        & room           & 10        \\ 
knowledge     & 21        & group        & 14        & together       & 10        \\ 
new           & 21        & present      & 14        &                &           \\ \hline
\end{tabular}
\caption{Frequency of terms found in abstracts after removing stop words.}
\label{tab:wordfrequencies}
\end{table}

As an initial index to recurring themes beyond the chosen search terms used in our query (RQ1), we present a word-cloud (see Figure \ref{fig:wordcloud}) and table depicting the frequency of the most common words in the publications' abstracts (see Table \ref{tab:wordfrequencies}). The frequency of particular keywords used in the query, while being a common denominator for all publications, shows which of the keywords were returned most often. For example, \textit{`communication'}, \textit{`remote'}, \textit{`video'}, and \textit{`teams'} were the four most often used terms. Particularly frequently used terms that were \textit{not} part of our query are: \textit{`research'}, \textit{`design'}, \textit{`paper'}, \textit{`participants'}, \textit{`use'}, \textit{`study'}, \textit{`different'}, and \textit{`people'}. This set of terms can be interpreted (as one of several possible interpretations) as follows: in the corpus of relevant publications, it is often central to \textit{research} and \textit{study} \textit{different} \textit{people} as \textit{participants} engaged in \textit{remote}, \textit{distributed} \textit{collaboration}. Overall, this can be seen as a general first level answer to RQ3 (topics covered in HCM) because this characterization seems representative of most of the publications and indicates a certain human-centeredness of the approaches. However, the interpretation also hints at a perceived emphasis on `remote' (45 mentions) collaboration in the publications, while for example the term `collocated' has 19 mentions, with the added variation of `co-located' with 5 mentions achieves only slightly more than half of the mentions of `remote' (24 vs. 45) or exactly two thirds when compared to `distributed' (24 vs. 36). 

Concerning our RQ2, the term `hybrid' is not in the list of the most frequently used terms (the list comprises words with at least ten mentions) with just five mentions in the abstracts, therefore, making it barely visible in the word-cloud (it is situated right below the first `e' in remote). This shows that the particular term `hybrid' was used only rarely in the years up until 2019 to characterize research described now under this notion. 

\subsection{Terms Used to Describe the HCM Phenomena in Papers}
%\textcolor{brown}{Make a chart/diagram showing how the authors referred to the systems, by having a look at the abstracts, Banu even and odd years, }

In addition to showing the word-cloud with the most frequently used words in the publications' abstracts, we investigated how authors referred to the phenomena of hybrid collaboration and hybrid meetings in their papers (RQ2). In order to do that, we went through the titles, abstracts, and keywords of these papers and picked the relevant terms referring to the communication styles, types of participants and modes of participation aimed at describing the ``remoteness'' aspect of the research. We counted the overall number of these terms and listed them from higher to lower frequencies of usage. 

Among our final set of 62 publications in the ACM DL, which include in their studies fully or partially hybrid meetings and collaboration supposedly, we found out that only one paper \cite{xu_attention_2017} from 2017 used the term `hybrid meetings' to refer to the phenomenon. Similarly, there is only one publication from 2018 \cite{neumayr2018}, which is co-authored by two of our co-authors, using the term `hybrid collaboration'. You can see the frequency of the rest of the terms used in the abstracts in \autoref{tab:terms}.

In order to get a better picture of the terminology used by those papers to describe the remoteness in their research, we find it important to have a look at the total count of similar keywords as well. For instance, for some of those terms such as `video mediated communication' and `video-mediated communication', where both ways of writing refer to the same concept, we can sum their numbers in total. When we consider the total count of similar keywords, the most frequently used umbrella term for describing remote working groups is `partially distributed teams' (16) including `partially distributed teams' (6), `distributed teams' (2), `partially distributed conceptual design teams' (1), `partially-distributed groups' (1), `partially-distributed work groups' (1), `distributed (virtual) teams' (1), `geographically distributed teams' (1), `geographically distributed work teams' (1), `distributed development teams' (1) and `temporally distributed teams' (1). In referring to the phenomenon of remote communication, the most frequently used term is `video mediated communication' (13) including `video mediated communication' (8), `video-mediated communication' (4) and `video mediated group communication' (1). 

\autoref{tab:terms} shows that there have been 91 different terms or keywords used in 62 papers on HCM and 82 of them are used only once. This means that the terminological diversity on referring to hybrid meetings and collaboration in the field of HCI is so large that even in the same paper different terms or keywords can be used to describe the same phenomena. This also lets some of the related work go unnoticed when the keywords in the literature search do not cover all these different terms. By showing such conceptual confusion, we hope to underline the need for having a shared understanding of these terms and the necessity for a unifying term.

% \begin{figure}
%   \includegraphics[width=\linewidth]{figures/Terms}
%   \caption{Terms used to describe research we regarded as hybrid and the terms' frequencies as used in the abstracts. \textcolor{red}{TODO: change to tabular when all is agreed upon.}}
%   \label{fig:terms}
% \end{figure}

\begin{table}[]
\caption{Terms used to describe research we regarded as hybrid and the terms' frequencies as used in the abstracts.}
\label{tab:terms}
\resizebox{\textwidth}{!}{%
\begin{tabular}{llllll}
\textbf{Term Used in the Abstract}               & \textbf{Frequency} & \textbf{Term Used in the Abstract}                     & \textbf{Frequency} & \textbf{Term Used in the Abstract}    & \textbf{Frequency}   \\ \hline
Video mediated communication                     & 9                  & Globally collaborative contexts                        & 1                  & Remote participation                  & 1                    \\
Partially distributed teams                      & 6                  & Group to group collaboration                           & 1                  & Remote players                        & 1                    \\
Remote participants                              & 4                  & Group-to-group collaboration                           & 1                  & Remotely located participants         & 1                    \\
Video-mediated communication                     & 4                  & Group-to-group videoconferencing                      & 1                  & Robot mediated communication          & 1                    \\
Virtual teams                                    & 4                  & Home video communication                               & 1                  & Robot-mediated communication          & 1                    \\
Computer-mediated communication                  & 3                  & Hybrid collaboration                                   & 1                  & Sympathetic remote collaboration      & 1                    \\
videoconferencing                               & 3                  & Hybrid meetings                                        & 1                  & Tabletop collaboration                & 1                    \\
Computer mediated communication                  & 2                  & Isolates                                               & 1                  & Telematic dining                      & 1                    \\
Distributed teams                                & 2                  & Local and remote participants                          & 1                  & Telepresence                          & 1                    \\
Asynchronous video                               & 1                  & Long-term video connections                            & 1                  & Temporally distributed teams          & 1                    \\
Blended interaction spaces                       & 1                  & Mediated parent-child contact                          & 1                  & Three way distributed collaboration   & 1                    \\
Collaborative research activities                & 1                  & Mixed reality collaborative environment                & 1                  & Video communication                   & 1                    \\
Collaborative Skypecasting                       & 1                  & Mobile video telephony                                 & 1                  & Video communication systems           & 1                    \\
Collaborative work                               & 1                  & Multimedia conferencing system                         & 1                  & Video conference                      & 1                    \\
Collaborative workspace system                   & 1                  & Multiscale communication                               & 1                  & Video conferencing                    & 1                    \\
Combined distance and on campus modes            & 1                  & Online meetings                                        & 1                  & Video mediated group communication    & 1                    \\
Computer-mediated collaboration                  & 1                  & Partially distributed conceptual design teams          & 1                  & Video streams                         & 1                    \\
Computer-mediated communication and presence     & 1                  & Partially-distributed groups                           & 1                  & Video-chat systems                    & 1                    \\
Computer-supported collaborative learning (CSCL) & 1                  & Partially-distributed work groups                      & 1                  & Video-mediated collaborative settings & 1                    \\
Distance separated participants                  & 1                  & Peripheral participants                                & 1                  & Video-mediated interaction            & 1                    \\
Distant colleagues                               & 1                  & Physically distant and computer mediated communication & 1                  & Video-mediated meetings               & 1                    \\
Distributed (virtual) teams                      & 1                  & Real-time, distributed conferences                     & 1                  & Virtual collaboration                 & 1                    \\
Distributed awareness                            & 1                  & Remote assistance system                               & 1                  & Virtual collaborative courses         & 1                    \\
Distributed design collaboration                 & 1                  & Remote collaborative physical tasks                    & 1                  & Virtual direct communication          & 1                    \\
Distributed development teams                    & 1                  & Remote collaborator                                    & 1                  & Virtual meeting room                  & 1                    \\
Distributed tabletop collaboration               & 1                  & Remote collaborators                                   & 1                  & Virtual meetings                      & 1                    \\
Distributed tangible environments                & 1                  & Remote communication                                   & 1                  & Virtual tabletop                      & 1                    \\
Distributed team collaboration                   & 1                  & Remote expertise                                       & 1                  & Virtually collocated teams            & 1                    \\
Distributed workspace                            & 1                  & Remote groups                                          & 1                  & Workplace communication               & 1                    \\
Geographically distributed teams                 & 1                  & Remote interaction                                     & 1                  &                                       & \multicolumn{1}{l}{} \\
Geographically distributed work teams            & 1                  & Remote meetings                                        & 1                  &                                       & \multicolumn{1}{l}{} \\ \hline
\end{tabular}}
\end{table}

\subsection{Citations and Venues}

From the 62 publications in the final set, eight were from journals and 54 from conference proceedings. The most prominent conference venue is the ACM Conference on Human Factors in Computing Systems (CHI) with 17 papers, followed by the ACM Conference on Computer-Supported Cooperative Work and Social Computing (CSCW) with 14 papers (not including one CSCW paper that was published in the Proceedings of the ACM on HCI (CSCW) because CSCW --- like many other conferences --- recently switched to a journal publishing method). The ACM International Conference on Supporting Group Work (GROUP) with four and the ACM Designing Interactive Systems (DIS) conference with two publications are the only \tn{further} sources of conference proceedings with more than one publication. Among the journals, only ACM Transactions on Computer-Human Interaction (TOCHI) is recurring with two publications.

\begin{figure}
  \includegraphics[width=\linewidth]{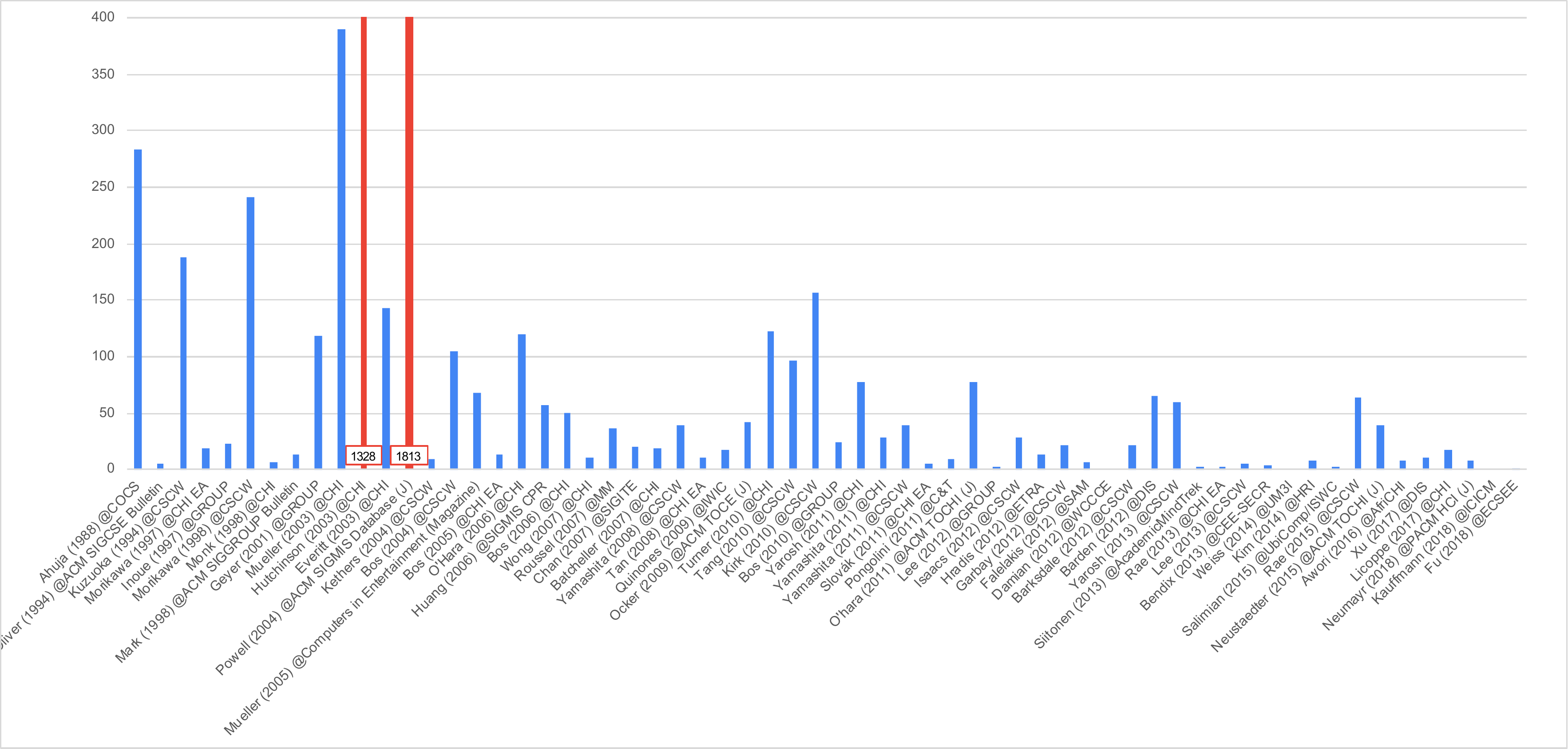}
  \caption{The citations of all publications as retrieved from Google Scholar on September 14, 2020.}
  \label{fig:citations}
\end{figure}

For an overview of the citation frequency of the publications, please refer to Figure \ref{fig:citations}, where also the short names of the publication venues are depicted next to the year and last name of the first author (further adding to RQ9). Two of the publications really stand out when it comes to citation frequency because they have 1,813 \cite{powell_virtual_2004} and 1,328 \cite{hutchinson_technology_2003} citations (which is indicated by red-colored bars and a label on the respective bar in Figure \ref{fig:citations}). %We decided against using a logarithmic scale because (1) the publications with zero citations were not displayable (as the logarithm of 0 is undefined) and (2) the quantitative relations between the publications would have been lost. 
The next most cited paper has the comparatively lower number of 390 citations \cite{mueller_exertion_2003}, and then only nine publications have more than 100 citations. The mean number of citations in our final corpus is 100.37 and the median is 21.5 (due to a lower number of crass outliers as mentioned before) which shows that it is a well cited selection. Recently, Correia et al. \cite{correia2018} identified the average number of citations in the broader field of CSCW as measured through the venues JCSCW, ACM CSCW, GROUP, and ECSCW to be at least 39 per paper, and therefore, vaguely comparable to our subset of publications on HCM. Only four of the publications in our corpus have either zero citations or one citation (from the years 2012, 2014 and two of 2018 which still have the potential to receive citations in the upcoming years) which together with the average  number of citations hints at a reasonable impact of the published research. However, when looking at the chart, one could gain the impression that the most influential publications were created starting at the end of the 1980s and up until the first half of the 2000s. Part of this phenomenon can be explained by an ambivalence that usually older publications have more citations than more recent ones because there simply were more opportunities to cite them which is however not always true in the field of CSCW, where there is a trend for citing more recent papers more often \cite{horn2004}. The historical situation concerning the number of citations seems to be, however, in line with the general interest in the topic of CSCW, e.g. the ACM CSCW conference was established in 1986, and since then some waves of interest can be observed. Additionally, the topics established in this regard are now more or less part of many different domains and on their way to becoming general knowledge in these domains. For example, a software developer is almost unavoidably concerned with some functionality initially inherent only to classical groupware when creating a smart phone application.

\subsection{Authors}
\label{sec:authors}
For our research questions, mainly regarding RQ1, it is interesting to see how many of the top authors in the field of CSCW also published research concerned with HCM. From the 57 authors with more than ten publications in the broader field of CSCW as identified by Correia et al. \cite{correia2018}, seven also published with direct relevance to hybrid collaboration and hybrid meetings (not necessarily as first authors). For an overview, please see Table \ref{tab:authors}. 

%\textcolor{red}{(TODO: Add some thoughts about the most cited authors in our corpus??.)}

\begin{table}
\centering
\caption{Most important CSCW authors' (as identified in 2016 by \cite{correia2018}) research with direct relevance to HCM (2019).}
\begin{tabular}{lllll}
\textbf{Author}  & \textbf{\# Pub. (CSCW)} & \textbf{\# Cit. (CSCW)} & \textbf{\# Pub. (HCM)} & \textbf{\# Cit. (HCM)}  \\
\hline
Gloria Mark      & 20                      & 696                     & 1                         & 14                         \\
Kori M. Inkpen   & 20                      & 847                     & 2                         & 118                        \\
Saul Greenberg   & 19                      & 1893                    & 1                         & 40                         \\
Susan R. Fussell & 17                      & 1368                    & 2                         & 29                         \\
Mary Beth Rosson & 14                      & 1434                    & 1                         & 42                         \\
Werner Geyer     & 11                      & 230                     & 1                         & 118                        \\
John C. Tang     & 10                      & 168                     & 2                         & 161   \\  
\hline
\end{tabular}
\label{tab:authors}
\end{table}

From these seven authors active in the sub-field of HCM, only three (John C. Tang, Werner Geyer, and to a lesser extent Kori M. Inkpen) received a considerable share of their citations in this sub field. We do not provide further analysis (e.g., percentages) because it is illegitimate to compare the citations reported by Correia et al. \cite{correia2018} between May and July of 2016 and those in our own research extracted more than three years later (in November, 2019), while both approaches at least used the same data source (i.e., Google Scholar). Instead, we aim at giving a quick estimate of how magnitudes between the two sets, i.e. CSCW in general as the superset and HCM as the subset, are characterized when it comes to the most active and cited authors. Consequently, we feel that there is a sufficient overlap between authors prominent in the broader field of CSCW and those active in HCM to show that this subset is well rooted in CSCW. 
That being said, with the exceptions of Saul Greenberg and John C. Tang, this author list excludes researchers associated with Media Space research (to be discussed later), notably Robert Stults, pioneer of the concept, and Steven Harrison, its chief champion, and many others who published repeatedly in the area such as Victoria Bellotti, Sara Bly, Paul Dourish, Ellen Issacs, William Buxton, Christian Heath, Austin Henderson, Susan Irwin, William W. Gaver, Paul Luff, Marilyn Mantei, Carman Neustaedter, and Abigail Sellen. Had Media Spaces been included in the SLR, the list would have a very different composition. 

\subsection{Domains}
Concerning the different domains dealt with in the publications (see Table \ref{tab:domains}), we had a look at what the authors explicitly stated where they see their works settled, or, if no such information was present we deduced from the experiments and descriptions which domains could profit from the findings (RQ8). Please note that several publications were tagged with multiple domain codes. Among 62 publications, 14 of them were tagged with two domain codes and five of them were tagged with three domain codes. Overall, we see a focus on workplace settings (33), but also many papers are dealing with general findings (27) that can be transferred to a number of different application scenarios.

There have been outlier examples, which differ topic-wise from the rest of the publications. One of those examples is a  paper from 2016 by Awori et al. with the title ``Sessions with Grandma: Fostering Indigenous Knowledge Through Video Mediated Communication'' tagged as the domain code of ``general'' and the paper is about the usage of video-mediated communication for sharing indigenous knowledge between elderly people located in rural areas and diaspora youth living in cosmopolitan cities of Kenya \cite{awori_sessions_2016}. Another interesting example is ``Telematic Dinner Party: Designing for Togetherness Through Play and Performance'' from 2012 by Barden et al. \cite{barden_telematic_2012}, which focuses on an exploratory user study of a telematic system to support dinner parties among co-located and remote participants. This paper was tagged as ``domestic'', ``leisure'', and ``entertainment''. A publication from 2007 with the title ``Testing the Technology: Playing Games with Video Conferencing'' by Batcheller et al.  \cite{batcheller_testing_2007} was tagged as ``entertainment'', ``workplace'', and ``general'' and the paper discovers the gaming experience in a video-based setting through a questionnaire as well as direct observation of the users.

% \begin{figure}
%   \includegraphics[width=0.7\linewidth]{figures/domains}
%   \caption{Number of publications per domain. Please note that multiple domains may apply to a publication. \textcolor{red}{Maybe delete this chart.}}
%   \label{fig:domains}
% \end{figure}

\begin{table}[ht!]
\caption{Mapping between the domains and publications. Please note that multiple domains may apply to a publication. }
\label{tab:domains}
\begin{tabular}{ll}
 \textbf{Domain} & \textbf{Publications} \\
 \hline
 Workplace (33) & \cite{kauffmann_knowledge_2018, neumayr2018, licoppe_showing_2017, bendix_role_2013, lee_input-process-output_2013, siitonen_i_2013, barksdale_video_2012, garbay_normative_2012, lee_knowledge_2012, ohara_blended_2011, pongolini_global_2011, bos_shared_2010, tang_threes_2010, turner_exploring_2010, quinones_bridging_2009, tan_barriers_2008} \\
  & \cite{yamashita_impact_2008, batcheller_testing_2007, chan_facilitating_2007, wong_sharing_2007, bos_collocation_2006, huang_preliminary_2006, bos_traveling_2005, bos_-group/out-group_2004, kethers_remote_2004, powell_virtual_2004, everitt_two_2003, geyer_team_2001, mark_building_1998, inoue_integration_1997, morikawa_hypermirror:_1997, kuzuoka_gesturecam:_1994, ahuja_rapport_1988} \\
 General (27) & \cite{neumayr2018, xu_attention_2017, awori_sessions_2016, salimian_exploring_2015, weiss_models_2014, rae_using_2013, falelakis_automatic_2012, hradis_voice_2012, isaacs_integrating_2012, lee_knowledge_2012, ohara_blended_2011, slovak_exploring_2011, yamashita_improving_2011, yamashita_supporting_2011, batcheller_testing_2007, roussel_beyond_2007, huang_preliminary_2006} \\
  & \cite{ohara_everyday_2006, bos_traveling_2005, powell_virtual_2004, everitt_two_2003, geyer_team_2001, monk_peripheral_1998, morikawa_hypermirror:_1998, inoue_integration_1997, morikawa_hypermirror:_1997, ahuja_rapport_1988} \\
 Education (11) & \cite{fu_teaching_2018, xu_attention_2017, kim_is_2014, bendix_role_2013, damian_instructional_2012, lee_knowledge_2012, ocker_training_2009, quinones_bridging_2009, chan_facilitating_2007, huang_preliminary_2006, oliver_software_1994} \\
 Domestic (9) & \cite{licoppe_showing_2017, neustaedter_sharing_2015, rae_framework_2015, yarosh_almost_2013, barden_telematic_2012, yarosh_mediated_2011, kirk_home_2010, ohara_everyday_2006, hutchinson_technology_2003} \\
 Entertainment (4) & \cite{barden_telematic_2012, batcheller_testing_2007, mueller_sports_2005, mueller_exertion_2003} \\
 Leisure (3) & \cite{rae_framework_2015, barden_telematic_2012, isaacs_integrating_2012} \\
 Personal (2) & \cite{isaacs_integrating_2012, ohara_everyday_2006} \\
 \hline
\end{tabular}
\end{table}

\subsection{Study Types \& Data Collection}
%\textcolor{brown}{Give an overview about E/NE/RW..., participants, setting, data collection.}
%\textcolor{brown}{Create Matrix or describe which study types where used for publications at which conferences or in which articles.}
%\textcolor{brown}{Show year vs. changes in study types and experiment types.}

We tagged papers based on the type of the study conducted as `experiment', `naturalistic experiment' or `real world study' (RQ5). With the `experiment' tag, we refer to any type of controlled experiment, whereas with the `naturalistic experiment' tag, we differentiate publications, which conduct experiments imitating the HCM settings in the real world to a considerable extent. We also tagged ethnographic/field studies taking place in the actual world as `real world study'. Among 62 papers, only 15 papers (less than 25 percent) are purely real world studies. 21 (one third) of these papers are based only on controlled experiments and 13 of the papers are based on naturalistic experiments. This shows that more than half of the papers (34 out of 62) are experiment-based only. Apart from these numbers, one paper was a combined real world study and an experiment, another one was a mixture of real world study and a naturalistic experiment, and one paper was a literature review, which does not fit to any of these three categories. Nine papers were tagged as `not applicable' meaning that the type of the study is unknown or not explained in detail.

We also categorized the papers based on whether it was explicit or implicit in the paper that the study had a hybrid setting (RQ7). In other words, we wanted to know how much researchers were consciously including the hybrid setting in their studies. Among the 62 papers, 26 of them used the hybrid setting in their study explicitly (i.e., consciously or on purpose, but still without necessarily naming it `hybrid') whereas 28 of the papers included the hybrid setting implicitly (e.g., it was only a natural by-product of the settings) in their studies. One of the papers had both explicit and implicit cases of HCM and seven of the papers are marked as `not applicable'' for reasons of either the paper does not involve empirical research or we are not knowledgeable about the further details of their choice of setting.

Regarding the data collection, we can claim that many of the papers collected multiple types of data and merged and analyzed those data together. Only six of the papers did not include any empirical work. Out of 62 papers, 21 have collected individual-based interview data, whereas two of the papers have collected group interview data. Interviews are followed by questionnaires (20), video recordings (16), surveys (7), system logging/log files (7), observation (5), literature review (3), field notes (3), diaries/journals (3), group discussion/focus groups (2), performance (2), reflections (1), gaze recordings (1) and a rich case study data (1) consisting of six different cases and scenarios of challenges of distributed teams. While questionnaires and/or surveys were used in a large number of papers to gather data, in sum, only three papers are based on fully quantitative findings \cite{falelakis_automatic_2012, chan_facilitating_2007, monk_peripheral_1998}.

\section{Taxonomy of HCM}
\label{sec:taxonomy}

\begin{figure}
  \includegraphics[width=1\linewidth]{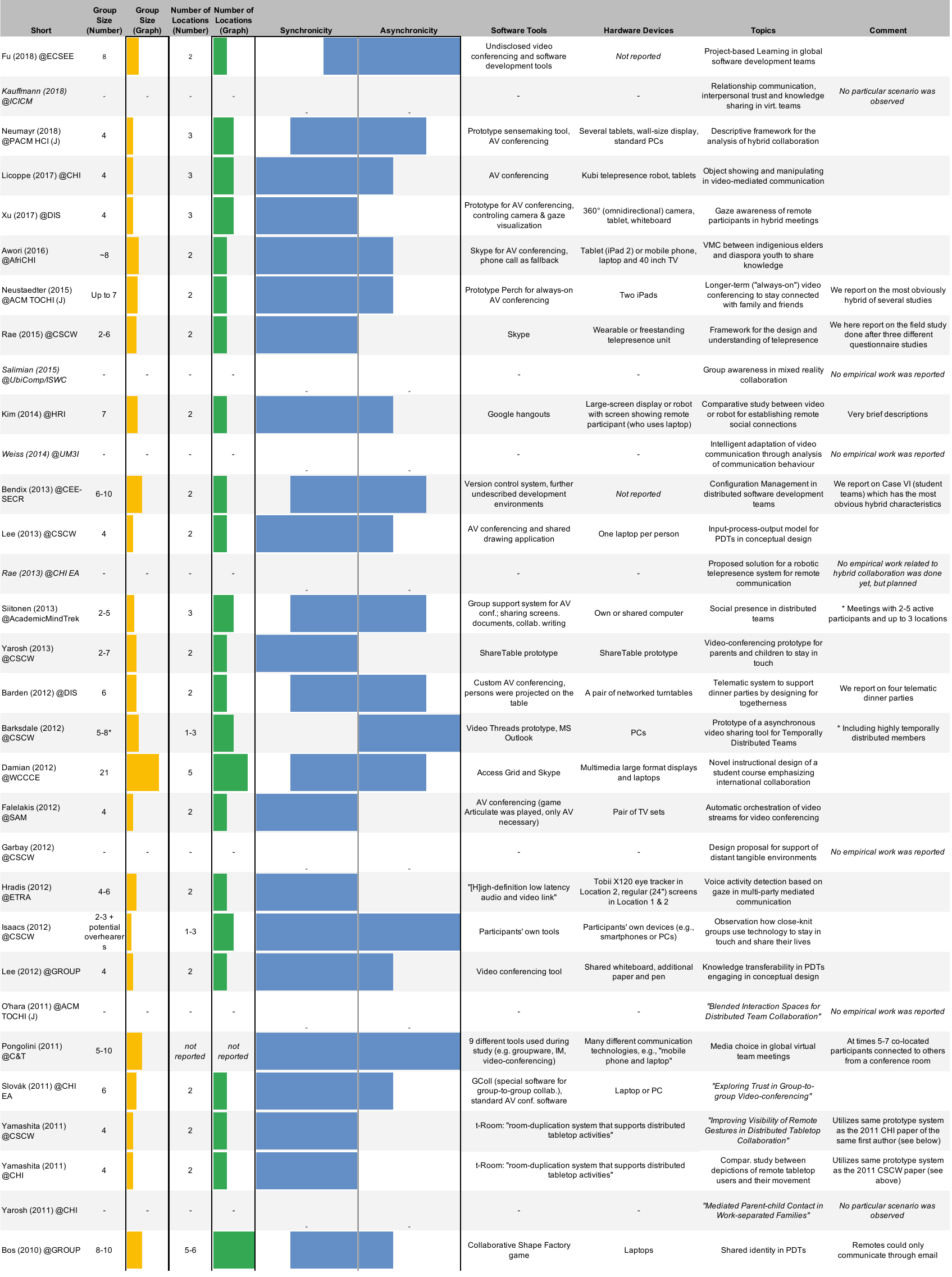}
  \caption{The taxonomy of HCM (Part 1 of 2).}
  \label{fig:taxonomy}
\end{figure}

\begin{figure}
  \includegraphics[width=1\linewidth]{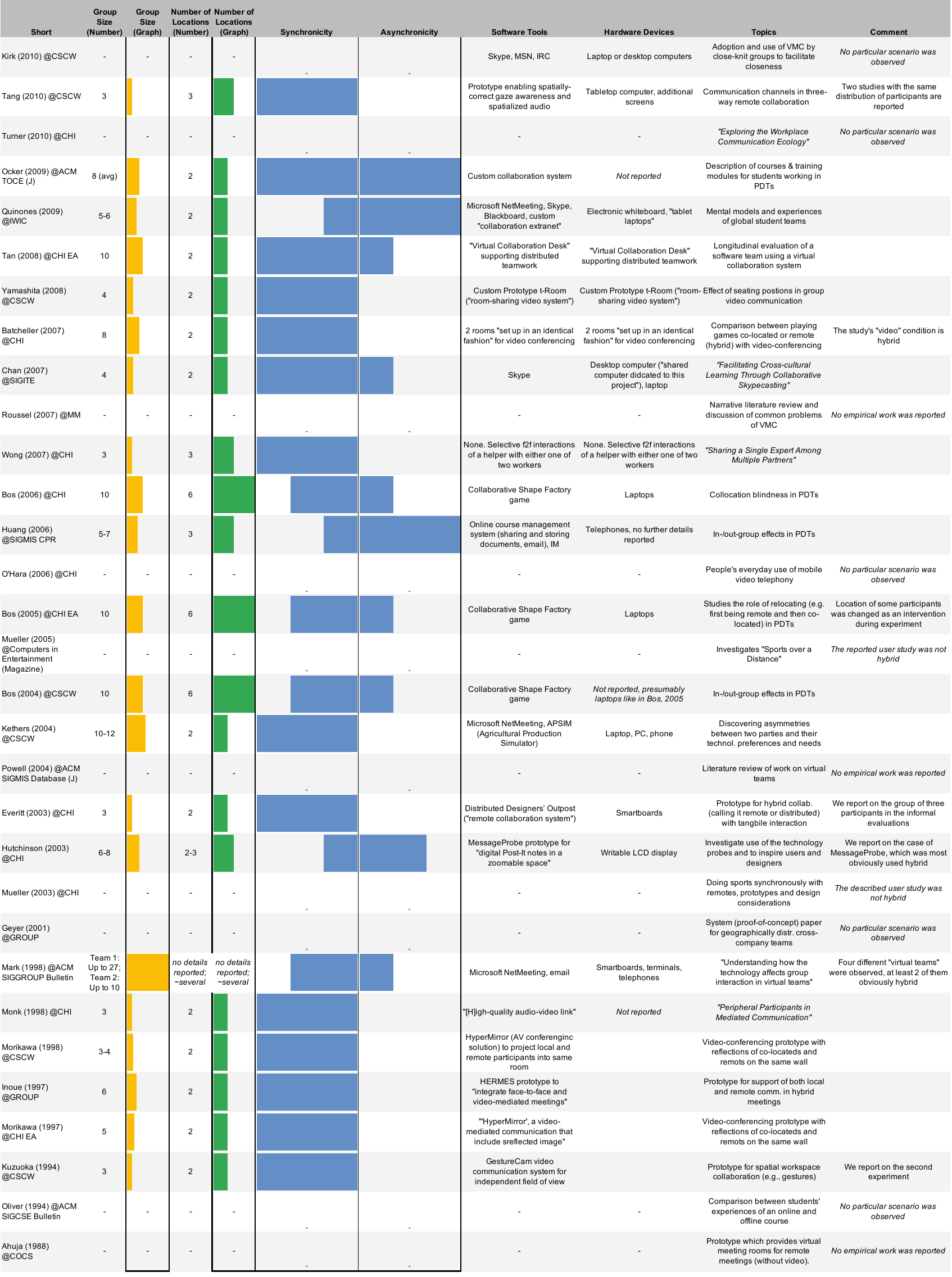}
  \caption{The taxonomy of HCM (Part 2 of 2).}
  \label{fig:taxonomy2}
\end{figure}
Based on the systematic literature review, we introduce a taxonomy for the description and categorization of HCM. The summarized findings of all 62 articles used to develop this taxonomy are shown at the end of Part A in \autoref{tab:findings}.

As stated in Section \ref{sec:def_hybrid_collaboration}, the definition of hybrid collaboration by Neumayr et al. \cite{neumayr2018} has three different parts. The first is the time-space matrix' dimensions, the second is tool and device usage, and the third is teams split into subgroups. The first two parts are thought to be basic and more-or-less obligatory prerequisites for the concept of hybrid collaboration while the subgroups part is comparatively optional for hybrid collaboration and even more so for hybrid meetings. Although we initially intended to show also the teams' behavior towards forming subgroups, this was not consistently possible because reports on subgroup formation only rarely occurred in the publications. Therefore, in the taxonomy, we show the more obligatory dimensions, that is, participants' location, synchronicity, and software tool \& hardware device usage (see Figures \ref{fig:taxonomy} and \ref{fig:taxonomy2}). By showing all these different information in the form of a taxonomy, we aim at answering RQ10 by suggesting a way for future researchers to categorize and thematically cluster the work on HCM beyond showing the mere results.

In the case of a publication relying on (prior) naturally occurring collaborative activities of the respondents (and did not observe a particular setting) in its empirical work (e.g., in the form of a questionnaire study) or where empirical work is altogether absent, several or all dimensions of the taxonomy might not be applicable and accordingly marked with a dash (-). Additionally, for such publications an according comment is available in the column `Comment', either as `No particular scenario was observed' or `No empirical work was reported'. The taxonomy's first column is reserved for a short name of the publication which is intended to be a unique identifier as well as giving some information about the first author, the year of the publication and the short name of the venue (or journal indicated with a (J)) where it was published. 
The following sections give details about how the taxonomy can be interpreted concerning the time-space matrix' dimensions in columns 2-7 (Section \ref{sec:taxonomy_timespace}), the software tool \& hardware device usage in columns 8 and 9 (Section \ref{sec:taxonomy_toolanddevice}), followed by an overview of how subgroups are discussed in related publications in Section \ref{sec:taxonomy_subgroups}, and, finally, meaningful examples in Section \ref{sec:meaningful_examples}.

\subsection{Time-Space Matrix' Dimensions}
\label{sec:taxonomy_timespace}
In order to cope with the vast breadth of the scenarios covered in this article, we consider the actual tasks and activities undertaken in the publications' studies, rather than the functionality of the used tools and devices. This means that for a publication describing the usage of a videoconferencing tool %(e.g., Zoom, Microsoft Teams, or Skype)
that is in principle capable of affording both synchronous and asynchronous interaction of both co-located and remote users, we aim at giving an account of how the tool was actually used in the studies. For example, if Skype was used in a user study by three co-located users for a brainwriting activity, the scenario is categorized as co-located and mainly synchronous. %Similarly, if the information given in the publication is sufficient to report on details of how the teams typically split into subgroups during the collaboration, we also estimate this according to the user study and task descriptions. In the example above, the brainstorming will most probably be done in a plenary session of all three participants, while a higher number of participants in combination with the task might lead to the teams splitting up into several subgroups working on subproblems of the overall task. This is especially true for partially distributed teams (and, therefore, teams engaging in HCM) that have a tendency to ``form strong `subgroups' rather than one cohesive group" according to Bos et al. \cite{bos_shared_2010}. 
One of the main authors went through the descriptions and extracted the information concerning the studies' time and space dimensions and then rated them according to the descriptions below. Ambiguous cases were thoroughly discussed among the two main authors. Still, some of the columns in the taxonomy are interpretative (mainly concerning synchronicity/asynchronicity) and should be seen as rough indicators to get an overview rather than as definitive classifications.
The following paragraphs detail on our understanding of the different parts of the definition as described in the taxonomy.

\subsubsection{Time}
Inspired by Lee and Paine \cite{lee15} the taxonomy understands Johansen's time-space matrix' dimension of time \cite{johansen_1988} as a continuum instead of a dichotomy. However, different to the approach in \cite{neumayr2018}, the taxonomy regards both `Synchronicity' as well as `Asynchronicity' as distinct continua, which paints a more truthful picture of the different scenarios and tasks that were described throughout the publications. For example, if a system affords both synchronous videoconferencing and the exchange of asynchronous messages (e.g., \cite{siitonen_i_2013,pongolini_global_2011}), and all of this functionality was used by the participants in the reported publication, both columns 
`Synchronicity' and `Asynchronicity' received high ratings. The ratings are depicted as bars representing values between zero (i.e., absent bar) and three (i.e., bar is fully filled) and the bars are oriented from right to left for column `Synchronicity' and left to right for column `Asynchronicity'. This results in a view similar to a scale to visualize if there was a focus on either synchronous or asynchronous interactions, i.e., whether there was more weight on either side of the scale.

We treated  \textit{synchronous collaboration} as one with only a few seconds between user interactions (which is typical for meeting-style collaboration) and \textit{asynchronous collaboration} as one with often several hours or days between user interactions. Additionally, in accordance to observations in \cite{neumayr2018}, current tool support for sharing artefacts prepared by collaborators in advance in hybrid settings is oftentimes weak. Considering the fictitious example that a co-located participant in a hybrid meeting  would like to show their weekly progress of an implementation task and might find it difficult to share their code with other (e.g., remote) participants, when there is only standard videoconferencing equipment in a meeting room, such as one camera and the presentation computer, which could be used for screen sharing, is controlled by someone else. Thus, we considered resources or artefacts prepared by collaborators (e.g., a document, presentation, or coded implementation) and brought into a collaboration session or meeting as a weak form as asynchronicity and accordingly categorized this as `1' in column 'Asynchronicity' (e.g., \cite{licoppe_showing_2017}). The complete absence of such asynchronously prepared material (in addition to the absence of other asynchronous interactions) is depicted in the taxonomy as `0' in column 'Asynchronicity' (in turn frequently leading to a categorization of `3' in column 'Synchronicity' for the purest form of synchronous collaboration, as in \cite{xu_attention_2017} or \cite{rae_framework_2015}). Likewise, rapid reciprocal user interactions within the bounds of merely a few seconds over the major part of the duration is depicted as `3' in column 'Synchronicity' while only having such behavior rarely leads to decreases in the ratings for synchronicity (e.g., \cite{slovak_exploring_2011,tang_threes_2010}).

%As a quick overview, we now report where the focus of the publications was concerning the time dimension. 
We counted the number of publications that had their focus in either one of synchronous or asynchronous interaction or where both played approx. the same role. Of the 45 publications we categorized in this regard, the majority of 32 (71.\.1\%) focused on synchronous interaction, only five (11.\.1\%) on asynchronous interaction and eight (17.\.7\%) describe relatively balanced settings with interaction behavior of both worlds.

\subsubsection{Space}
To give an account of the participants' physical distribution in the publications' studies concerning their individual `remoteness' status, the taxonomy shows the `Group Size' and the `Number of Locations'. Both attributes are depicted using a number (or number range) as well as a normalized bar visualization. The minimum and maximum values of these bars is determined by the lowest and highest value in this particular category (e.g., for `Group Size' the maximum is 27, for `Number of Locations' it is 6). Please note that the maximum values are global for the entire taxonomy and not only for any of the two parts. Please also note that for the visualization of number ranges, we used the maximum values (e.g. for a `Group Size' of `2-7', we chose to visualize the bar to represent the number `7' as this bar then also includes the smaller groups' sizes). The participants in these locations are treated as remote in relation to each other, meaning that only mediated communication is possible between locations, whereas direct in-person communication is possible within the locations. The taxonomy is agnostic towards possible hierarchies in the locations (e.g., whether there was a primary or control room or not), since this information was rarely provided, although we acknowledge that this could play an important role. %, however, we tended to present locations with a higher number of participants or better infrastructure, such as AV conferencing equipment, devices, or tools, in column `Location 1'.
28 (62.\.2\%) of the publications used two different locations, ten (22.\.2\%) used three (including three using up to three locations), five (11.\.2\%) used more than three locations, and two (4.\.4\%) publications did not disclose details about the number of locations. 

\subsection{Tool and Device Usage}
\label{sec:taxonomy_toolanddevice}
In the taxonomy's `Software Tools' and `Hardware Devices' columns we show that among the 62 papers, 16 did not mention any software tools or hardware devices. Seven out of these papers did not report any empirical work, whereas six out of these papers did not report any user study, and three of these papers were not based on a user study in a hybrid setting. Apart from these, two of the 62 papers did not specify their video conferencing and/or software development tools even though they used them in their studies.

Ten of the papers used at least one form of commercial software in their studies, the names of which are indicative of their age. Skype (6) \cite{awori_sessions_2016,rae_framework_2015,damian_instructional_2012,kirk_home_2010,quinones_bridging_2009, chan_facilitating_2007} stands out as the most used commercial software and is followed by Microsoft NetMeeting (3) \cite{quinones_bridging_2009,kethers_remote_2004,mark_building_1998}, Google Hangouts (1) \cite{kim_is_2014}, MS Outlook (1) \cite{barksdale_video_2012}, MSN (1) \cite{kirk_home_2010}, IRC (1) \cite{kirk_home_2010}, Blackboard (1) \cite{quinones_bridging_2009} and Access Grid (1) \cite{damian_instructional_2012}. While Skype, NetMeeting, Google Hangouts and Access Grid are used for audio/video (AV) conferencing, MS Outlook, MSN and IRC are used for text-based hybrid collaboration. One paper used a noncommercial software called APSIM (Agricultural Production Simulator) \cite{kethers_remote_2004} and at least one paper specified that they used an open source prototype for group-to-group collaboration called GColl \cite{slovak_exploring_2011}. Eleven of the papers used their own prototype specifically aimed for AV conferencing and 15 of the papers used their own prototype for supporting different forms of hybrid collaboration. 26 papers in total used at least one prototype for their HCM study. Ten papers used unnamed AV conferencing software.

Six of the papers, which used at least one software tool, did not report on any hardware devices used in their studies. While earlier studies (before 2000) used their own prototype as the hardware device as well, after 2000 we see a trend for the usage of more personal devices such as tablets and laptops. 17 papers used laptops or PCs in their studies, whereas five papers, which were published in 2015 and later, included tablets. 12 of the papers used at least one display, screen or TV set in their studies. Also two papers used room-to-room videoconferencing systems and a paper studied a hybrid meeting from a conference room. Two recent papers \cite{licoppe_showing_2017, rae_framework_2015} included telepresence robots and another recent paper \cite{kim_is_2014} had a robot with a screen showing the remote participant.  In terms of hybrid collaboration, at least seven of them used electronic/smart boards, three of them used tabletops \cite{yamashita_improving_2011, yamashita_supporting_2011, yarosh_mediated_2011}, and one paper used a pair of networked turntables \cite{barden_telematic_2012}. One paper claimed that they used nine different tools in their study without specifying any \cite{pongolini_global_2011}.

%For example, video- or audio-based and which tools/prototypes they used. Commercial tools vs. prototypes, hardware/software.

%\textcolor{red}{TODO:} XX of the publications used AV conferencing, XX relied on audio solely, XX used text-based chat. XX made some use of commercially available tools. The hardware devices comprised standard PCs, wall-size displays, tablets, wearables ... 

\subsection{Subgroups}
\label{sec:taxonomy_subgroups}
The definition of hybrid collaboration that we use~\cite{neumayr2018} states that teams typically split up into several subgroups. Although we do not see this as a strict criterion, it is common to find teams splitting into subgroups to do focused work and reforming a larger group to discuss results of the subgroup work, and repeating this process several times. It may be less common in some hybrid meetings %It is %certainly not a part of the definition
%not defining for hybrid meetings, because there oftentimes no splitting up into subgroups can be observed 
because ongoing group communication is a prerequisite for many meeting activities, but also because it has been technically difficult. First, audio separation of subgroup conversations in the local activity space such that remote participants can both hear and engage is a perennially thorny problem. Second, even in all-remote settings, sufficiently flexible AV transmission technologies to enable easy composition--decomposition--and re-composition in research prototypes have only been available since around 2014 (e.g. WebRTC).  Commercial products with `breakout' room capabilities  existed somewhat earlier, especially for webinars, but, again, these tended to focus on creating subgroups in all-remote contexts.

This is the most difficult dimension to extract from the descriptions in the publications because the details of the task descriptions (if present) are not always sufficient to deduce the typical number of subgroups and only few of the publications directly reported details on the teams splitting up into subgroups. %Nevertheless, we interpreted the descriptions and report on the typical number of subgroups. In cases, where no information was found or could be deduced, this column is filled with `not reported'. In ambiguous cases, 
Therefore, we do not report the details in the taxonomy, but give an overview of the papers that discuss the idea of subgroups below. With regards to future studies, we encourage the research community to include details about any observed behavior towards splitting up into subgroups. Such information could then be used to extend the taxonomy with a column `Subgroups' according to our initial plans. The cautious choice for readers is to assume a range of the theoretical minimum and maximum values, which are between 1 (the group collaborates as a whole), and the number of team members (when everyone currently contributes individually).

%The following descriptions are intended to exemplify how some of the publications dealt with the topic of subgroups. 

Neumayr et al. \cite{neumayr2018} includes descriptions of how the teams split up into subgroups at several occasions as this is one of the foci of the paper. All extreme values concerning the number of subgroups have been identified in their study, between 1 and 4 (number of team members). The authors suggested a measure for the team fragmentation (called the  ``Time-weighted Mean Number of Concurrent Subgroups (TNS)''), i.e., in how many subgroups a team splits up on average over the duration of the collaboration. The TNS numbers for their six teams have been between approx. 2 and 3 meaning that the teams on average split up into between 2 and 3 subgroups.

Some further publications' stance on subgroups is briefly summarized as follows:
\begin{itemize}
    \item Bos et al. \cite{bos_shared_2010} present some thoughts about the tendency of subgroups to form influenced by location, and how subgroups pertain to group identity in their related work.
    \item Ocker et al. \cite{ocker_training_2009} also cover subgroups in their literature review and use the concept of subgroups several times throughout the investigated constructs and items (but not when describing their results).
    \item Quinones et al. \cite{quinones_bridging_2009} only mention the existence of several subgroups insofar as they describe the parts of geographically distributed teams as subgroups of a team (e.g., ``Israeli subgroups'').
    \item Similarly, Bos et al. \cite{bos_traveling_2005} only mention subgroups when they are talking about parts of the group that were relocated (changes in the location from co-located to remote participants).
    \item Furthermore, Bos et al. in \cite{bos_-group/out-group_2004} describe the concept in their related work at some length and comment in the discussion that in their study in accordance to literature, subgroups formed ``based on location''.
    \item Finally, Geyer et al. \cite{geyer_team_2001} only mention one implicit comment about the existence of subgroups.
\end{itemize}

\subsection{Meaningful Examples}
\label{sec:meaningful_examples}
In this section, we present illustrative examples of how the taxonomy can be interpreted and justify our categorization choices. We describe the actual characteristics of the study settings or the envisioned usage scenarios concerning their hybrid features. Only a small number of publications particularly used the term `hybrid' to refer to their study settings or scenarios. As indicated above, we, nevertheless, also listed such publications that fulfilled our definitions (see Sections \ref{sec:def_hybrid_collaboration} and \ref{sec:def_hybrid_meetings}) regardless of the actual term usage. Most of the publications \textbf{i)} focused on a \textbf{particular set of settings} with hybrid characteristics (e.g., through empirical work in the form of experiments or user studies), while others \textbf{ii)} used empirical methods to learn from (prior) \textbf{naturally occurring collaborative activities} of the respondents, such as \cite{kauffmann_knowledge_2018,yarosh_mediated_2011,kirk_home_2010,turner_exploring_2010,ohara_everyday_2006,geyer_team_2001,oliver_software_1994}, and \textbf{iii)} a third category of the publications did \textbf{not report on empirical data} at all (e.g., descriptions of system implementation or design proposals) such as \cite{salimian_exploring_2015,weiss_models_2014,garbay_normative_2012,ohara_blended_2011,roussel_beyond_2007,powell_virtual_2004,ahuja_rapport_1988}. It was only possible to report all details of the different hybrid characteristics of the first category's publications in the taxonomy (\textbf{i}), that is where a relatively homogeneous design space of the studies was reported. Some of those studies are described and their categorization in the taxonomy is explained in Section \ref{sec:publications_focus_particular_setting}. For the latter two categories' publications (\textbf{ii} and \textbf{iii}) it was not possible to report details in the taxonomy because of the settings' heterogeneity (\textbf{ii}) or absence of studies (\textbf{iii}). Some examples for those publications are described in Sections \ref{sec:publications_focus_not_particular_setting} and \ref{sec:publications_no_empirical_work}, respectively.

\subsubsection{Publications Focusing on a Particular Hybrid Setting}
\label{sec:publications_focus_particular_setting}
We selected some illustrative examples of categorizations in our taxonomy that cover several different group sizes and numbers of locations as well as all levels (0-3) of `Synchronicity' and `Asynchronicity' to explain further details of our choices and present them in chronologically descending order.

Neumayr et al. \cite{neumayr2018} conducted a lab study with teams of four (see column `Group Size') where two co-located participants in one room and two further participants individually collaborated with them from separate rooms, resulting in three different locations. Only mediated communication was possible between the locations while in-person communication was possible within the locations (which is only useful for the location with two participants). The number of locations is indicated in column `Number of Locations' in the taxonomy. Besides an audio-video connection for synchronous communication, participants operated a custom-built prototype application where notes and likes could be left to be later discovered by others (or revisited by oneself) resulting in some form of asynchronicity. Additionally, a shared interactive whiteboard was part of the functionality that was used synchronously as well as asynchronously. The column `Synchronicity' was rated with 2 out of 3 points because the extent of the synchronous collaboration was not as high as in a discussion or meeting task but, nevertheless, reasonably high as compared to only rarely or never interacting synchronously. The column `Asynchronicity' was again rated with 2 out of 3 points because the functionality and usage regarding this (i.e., leaving notes, likes, or interpreting changes in the shared interactive whiteboard) was clearly above the mere bringing and sharing of prepared artefacts, which was commented above as a minimum requirement to receive a rating of 1 out of 3 points. Yet, scenarios in other publications had interactions of several hours or even days or weeks in between interactions, which is again clearly above what was described in \cite{neumayr2018} concerning asynchronicity. 

Licoppe et al. \cite{licoppe_showing_2017} describe a setting where a pair of co-located participants was connected to two separate remote parties in the second study they reported on. This leads to column `Group Size' filled out with 4 and `Number of Locations' filled out with 3. They focus on synchronous collaboration (column `Synchronicity' receives a rating of 3 out of 3) in the form of showing objects to remotes while the task also includes asynchronous artefacts in the form of artworks brought into the collaboration by the collocateds for the shared task of creating an art exhibition, hence, fulfilling the minimum requirement for some asynchronicity and a rating of 1 out of 3 in this column. The 'Software Tools' and 'Hardware Devices' are listed, consequently.

Xu et al. \cite{xu_attention_2017} conducted a lab study with teams of four (`Group Size' is 4), where two remotes were connected to two collocateds (`Number of Locations' is 3) seated in a conference room containing a prototype system consisting of a 360-degree camera and a screen indicating in simplified form where the remotes users' gazes were currently headed, that is, which portion of the individually adjustable 360-degree camera the remotes currently viewed. The publication focuses on synchronous interaction --- which is also emphasized through the publication's keywords --- in the form of a discussion task where no information about asynchronous resources was identified (`Synchronicity' is 3, `Asynchronicity' is 0).

Barksdale et al. \cite{barksdale_video_2012} describe a trial study of a prototype system for asynchronous communication (called Video Threads) that was deployed in three different locations (`Number of Locations' is 3) and used by between five and eight participants distributed around these locations ('Group Size' is 5-8, including highly temporally distributed members from different time zones). The prototype is intended to allow sending video messages instead of e-mail and having a strict focus on asynchronous interaction with typically hours or days (in the case of video messages received in after-work hours or during the weekend) between interactions due to time zone differences (`Synchronicity' is 0, `Asynchronicity' is 3).

Tang et al. \cite{tang_threes_2010} developed and evaluated a custom-built prototype for three-way collaboration, that is, a collaboration between three users that are remote to each other, leading to a reported `Group Size' of 3 and `Number of Locations' of 3. The strongly synchronous tasks of jointly re-arranging tiles on a shared interactive tabletop computer while an always-on audio-video link was active during all sessions in connection with the absence of any reports of asynchronous features led to the categorizations for `Synchronicity' as 3, and for `Asynchronicity' as 0. Please note that in principle, this example infringes the rather obligatory parts of the definition of hybrid collaboration because no collocateds were part of the setting. However, the actual setting provoked by the authors makes it appear for the remotes to be co-located with others, or at least having an impression of co-presence, as the authors state. The seating arrangement and having a separate screen and camera for each of the three participants, therefore, led after discussion, to an inclusion of this paper. 

Similarly, the publication by Wong et al. \cite{wong_sharing_2007} was the subject of discussion among the authors, because there, too, the definitions of hybrid collaboration concerning the dimension of space were initially thought to be infringed. Contrary to Tang et al.'s Three's Company \cite{tang_threes_2010} in Wong et al.'s publication, not collocateds are missing but remotes. However, the study was done to simulate remoteness by installing a partitioning wall between two of the participants and allowing a third participant to selectively engage in (in-person) co-located collaboration with either one of the two separated participants. Here, we decided to categorize the groups of three (`Group Size' is 3) as being distributed around three (simulated) locations (`Number of Locations' is 3). 

Quinones et al. \cite{quinones_bridging_2009} present a study in an educational context, where geographically distributed student teams work on construction and civil engineering tasks within a 16-week course. Teams consisted of five or six members (`Group Size' is 5-6) that were distributed between two locations (`Number of Locations' is 2). The teams at each location worked independently for most part of the collaboration but there were approx. weekly scheduled meetings during class with the addition of some meetings outside of class. We classified this relatively low level of `Synchronicity' as 1 and the focus on `Asynchronicity' as 3. 

\subsubsection{Publications Without an Identifiable Hybrid Setting}
\label{sec:publications_focus_not_particular_setting}
Overall, there are seven publications \cite{kauffmann_knowledge_2018,yarosh_mediated_2011,kirk_home_2010,turner_exploring_2010,ohara_everyday_2006,geyer_team_2001,oliver_software_1994} where no hybrid setting was identifiable from the descriptions. Most of these publications report on questionnaire studies, interviews, or researchers' self-reports of people engaging in a variety of naturally occurring HCM, therefore, making generalizations and classification in the taxonomy difficult.
For example, Kauffmann and Carmi \cite{kauffmann_knowledge_2018} describe a questionnaire study with 259 responses from all over the world engaging in a myriad of different settings. Therefore, it was not possible to give a truthful account of particular settings in the taxonomy. Another example is Yarosh et al. \cite{yarosh_mediated_2011}, where 14 parent-child pairs from work-separated families were interviewed about their strategies to stay in touch. Similarly, Kirk et al. \cite{kirk_home_2010} interviewed 17 participants spread across twelve different homes about their use of video-mediated communication. Furthermore, Turner et al. \cite{turner_exploring_2010} explored different communication technologies used in the workplaces of a small corporation by conducting two different surveys as well as more focused interviews of 23 employees over the period of one year. Moreover, O'Hara et al. \cite{ohara_everyday_2006} report on the naturally occurring mobile video telephony activities of 21 participants that were surveyed and interviewed. Geyer et al. \cite{geyer_team_2001}, then, describe a proof-of-concept for a system supporting collaborative activities (including hybrid settings) of geographically distributed cross-company teams and report on their own experiences with the system. Finally, Oliver \cite{oliver_software_1994} describes the results of a postal survey of students (67 replies) about their experiences with distance education in group work, hence, also covering a broad range of different settings. 

\subsubsection{Publications Without Empirical Work}
\label{sec:publications_no_empirical_work}
Overall, there are seven publications \cite{salimian_exploring_2015,weiss_models_2014,garbay_normative_2012,ohara_blended_2011,roussel_beyond_2007,powell_virtual_2004,ahuja_rapport_1988} that did not report any empirical work, rendering a classification in the taxonomy as impossible. Additionally, Rae \cite{rae_using_2013} conducted several studies with dyads in a non-hybrid setting but described concrete plans for future studies of hybrid settings. Similarly, Mueller \& Agamanolis \cite{mueller_sports_2005} and Mueller et al. \cite{mueller_exertion_2003} report several hybrid use cases of extertion interfaces, where several co-located players interact with a game at each location. However, the authors merely report on pairs of participants making use of their prototypes in user studies.
For all of these publications that describe concrete future plans for studying hybrid settings \cite{rae_using_2013}, or reflect on particular hybrid use cases of their proposed systems \cite{mueller_sports_2005,mueller_exertion_2003}, we abstained from doing a classification in our taxonomy due to the absence of hybrid empirical work. 

% \usepackage{graphicx}
% \usepackage[normalem]{ulem}
\begin{longtable}[c]{p{1cm}p{1.5cm}p{1cm}p{1cm}p{7.5cm}}
\caption{Summary of findings from the 62 included SLR publications  \label{tab:findings}}\\
% \resizebox{\textwidth}{!}{%
% \begin{tabular}{p{0.5cm}p{1.5cm}p{4.5cm}p{1cm}p{2cm}p{10cm}}
\toprule
\textbf{Year} & \textbf{Authors} & \textbf{Ref} & \textbf{FP} & \textbf{Findings}   \\ \midrule
\endfirsthead
\textbf{Year} & \textbf{Authors} & \textbf{Ref} & \textbf{FP} & \textbf{Findings} \\ \midrule
\endhead
\multicolumn{5}{ r }{\textit{(Continued)}}\\
\endfoot
\endlastfoot

2018                                          & Fu et al.                                                                                                                                                                          & \cite{fu_teaching_2018} & HIM                              & Co-located team members showed a higher level of group cohesion compared to the remote participants. Virtual communication %was sometimes found difficult due to 
lacked %of 
visual cues. %Bachelor's students 
Undergraduates were more frustrated %with virtual collaborative courses 
than %the master's 
graduate students.                                                                                                                                                                                                                                                                                                                                                                                                                                                            \\
2018                                          & Kauffmann \& Carmi                                                                                                                                                                                                                                                         & \cite{kauffmann_knowledge_2018} & SI                              & %The study found significantly 
Positive correlations between relationship communication, trust, and knowledge sharing. Interpersonal trust plays an important role in mediating the relationship between task communication and knowledge sharing. Only %the 
cognitive trust was found to %be a mediator for 
mediate knowledge sharing.                                                                                                                                                                                                                                                                                                                                                                                 \\
2018                                          & Neumayr et al.                                                                                                                                                                             & \cite{neumayr2018}    & HIM                              & Hybrid collaboration is investigated in-depth and different analytic tools are suggested %concerning the 
for exploring team fragmentation, subgroup configuration and volatility of groups.  %Furthermore, based on 
A novel notation system called "Domino", a more extensive description of hybrid collaboration, and nine different hybrid coupling styles are proposed.                                                                                                                                                                                                                                                                                                                                                                                                \\
2017                                          & Licoppe et al.                                                                                                                                                                 & \cite{licoppe_showing_2017}                & HIM                              & Focusing on showing objects with or without tools during domestic %the 
video-mediated communication% among family members and friends.
%this paper 
. Discusses showing objects only when relevant to the conversation, or when the showing process may take longer due to discussions or more detailed displays% regarding the particular object
. %Authors also list 
Suggests design implications to %better 
support %participants throughout the process of 
showing objects% during video-based meetings
.                                                                                                                                                                                                                    \\
2017                                          & Xu et al.                                                                                                                                                                                                                                                 & \cite{xu_attention_2017}           & HM                              & Simulated gazes increase feeling of presence, but do not always reflect what the remote participants pay attention to.                                                                                                                                                                                                                                                                                                                                                                                                                                                                                                                                                                                                                                  \\
2016                                          & Awori et al.                                                                                                                                                                                                                        & \cite{awori_sessions_2016}                        & HM                              & Managing video-conferencing during a moving classroom can be difficult for elderly people. Hands-free mobile tools are needed for a better live interaction.                                                                                                                                                                                                                                                                                                                                                                                                                                                                                                                                                                                            \\
2015                                          & Neustaedter et al.                                                                                                                            &                               \cite{neustaedter_sharing_2015}                                       & HM                              & Technologies supporting long-term domestic connections should move beyond conversations and focus on sharing everyday life. One-Size-Fits-All Solutions do not work and such technologies should run on dedicated devices.                                                                                                                                                                                                                                                                                                                                                                                                                                                                                                                              \\
2015                                          & Rae et al.                                                                                                                                                                                                                   &                             \cite{rae_framework_2015}                                         & TP                              & Proposes a framework for understanding telepresence including seven design dimensions% is proposed
, including the Initiation, Physical Environment, Mobility, Vision, Social Environment, Communication, and Interdependence% and is intended for use by designers as well as researchers
.                                                                                                                                                                                                                                                                                                                                                                                                                                                 \\
2015                                          & Salimian                                                                                                                                                                                                                                                              & \cite{salimian_exploring_2015}                              & HIM                              & %Researchers 
Describes a mixed reality collaborative environment %they developed 
involving a physical tabletop and virtual arm embodiments of the remote collaborator while interacting with the virtual tabletop% They mention their planned study and research questions briefly, no findings yet
.                                                                                                                                                                                                                                                                                                                                                                                                                                                       \\
2014                                          & Kim et al.                                                                                                                                                                                          & \cite{kim_is_2014} & TP-robot                        & Classroom participants showed more interest and empathy when they interacted with participants through Mobile Robotic telePresence than via traditional videoconference.                                                                                                                                                                                                                                                                                                                                                                                                                                                                                                                                                                                                                         \\
2014                                          & Weiss et al.                                                                                                                                                                                                         & \cite{weiss_models_2014}                              & HE                              & %The paper is about 
Reports an automatic real-time decision making system aimed for video-based communication and its implementation, which includes cue extraction, fusion and interpretation, and decision making processes to choose the best display of the videos.                                                                                                                                                                                                                                                                                                                                                                                                                                                                                          \\
2013                                          & Bendix \& Pendleton                                                                                                                                                                                                                                & \cite{bendix_role_2013}                                  & HT                              & %This paper brings into light 
Explores how distributed development teams can benefit from Configuration Management concepts. %Authors argue that 
Many problems %occuring 
in distributed development teams can be solved by applying the already existing solutions% in Configuration Management
.                                                                                                                                                                                                                                                                                                                                                                                                                                                                \\
2013                                          & Lee et al.                                                                                                                                                                                       & \cite{lee_input-process-output_2013}                                      & HT                              & %Based on a qualitative study, authors bring into light the 
Report the factors shaping the formation of shared understanding among partially distributed conceptual design teams. %In order to do that, they 
Present an input-process-output model: %as their result with 
inputs (personality, background/major, team formation, task, communication technology) influencing the process (factors contributing to shared understanding: awareness, interactivity, knowledge transferability, cohesiveness, trust, collaboration pattern, role emergence, conflict) leading to output (performance and communication satisfaction).                                                                                                                                    \\
2013                                          & Rae                                                                                                                                                                                                                                                                     & \cite{rae_using_2013}                                    & TP-robot                        & %This is a doctoral consortium paper presenting the thesis research plan 
Focuses on how important features of Mobile Robotic telePresence, such as physical embodiment and control over the system, %can 
lead to fruitful collaboration.                                                                                                                                                                                                                                                                                                                                                                                                                                                                                                                  \\
2013                                          & Siitonen \& Olbertz-Siitonen                                                                                                                                                                                                                  &               \cite{siitonen_i_2013}                                                        & HT                              & Based on real-world observations of distributed teams, researchers discuss how presence is socially constructed in interaction and underline the temporality and multispatiality of presence as well as its emergence as a strategic choice or as an example of playfulness during online collaboration.                                                                                                                                                                                                                                                                                                                                                                                                                                                 \\
2013                                          & Yarosh et al.                                                                                                                                                                                         & \cite{yarosh_almost_2013}                                         & HM                              & Setting up a tool called "ShareTable", which supports both video-chat and a tabletop task space, researchers observed two divorced families' usage% of the tool for four weeks each
. %They found that 
The tool enables a better sense of touch and closeness, however adjusting the frequency and density of the social contact with their kids can be a challenge for the divorced parents.                                                                                                                                                                                                                                                                                                                                                              \\
2012                                          & Barden et al.                                                                                                              & \cite{barden_telematic_2012}                                         & HM                              & Reports a telematic dinner party experience, bringing both remote and co-located participants around a table to eat together. Argues that cultural background of the participants and the social structure plays an important role in designing similar technologies.                                                                                                                                                                                                                                                                                                                                                                                                                                                                           \\
2012                                          & Barksdale et al.                                                                                                                                                        & \cite{barksdale_video_2012}                                         & HT                              & Reports %Researchers developed and deployed a tool called 
on "VideoThreads", which vizualizes video messages as a thread. %Studying 
Four teams %using this tool researchers argue that 
of participants enjoyed using VideoThreads %for general talk and conversations helping them get 
because it helped them become familiar with each other. 52\% of the participants found it more useful %or much more useful 
than email, but %However, 
others found %the tool 
it less useful than email %since 
because editing and rerecording videos, as well as searching through the video and audio content, is not %that 
easy.                                                                                                                                                                                                  \\
2012                                          & Damian et al.                                                                                                                                                                     & \cite{damian_instructional_2012}                                         & HIM                              & %This paper is about 
Reports on a 12-week-long graduate-level course in Canada involving international collaboration with foreign students connecting remotely from the UK. %Authors 
Suggests %strategies and insights throughout the course and suggest 
that international collaboration with students from different contexts and using computer-mediated communication in distributed collaboration improve %the 
learning experiences and broadens students' horizons.                                                                                                                                                                                                                                                                                          \\
2012                                          & Falelakis et al.                                                                                                                                                                                   & \cite{falelakis_automatic_2012}                                         & Heco                            & %This paper 
Focuses on the effects of automatic orchestration, which refers to providing intelligent selection of the most efficient camera views during a video-based meeting. Authors studied this with two distributed groups playing a collaborative board game and found that orchestration is useful similar to the live video-mixing done by the human editors.                                                                                                                                                                                                                                                                                                                                                                                   \\
2012                                          & Garbay et al.                                                                                                                                                                                            & \cite{garbay_normative_2012}                                        & Heco                            & %Authors share their 
Presents a design idea within the framework of activity theory, supporting distant tangible collaboration through a normative multi-agent, trace-based approach.                                                                                                                                                                                                                                                                                                                                                                                                                                                                                                                                                                               \\
2012                                          & Hradis et al.                                                                                                                                                                                                               &                        \cite{hradis_voice_2012}                                               & HM                              & Using statistical machine learning methods, %researchers 
reports experiments on whether it is possible to detect the active speaker in video-mediated communication based on the gaze and eye movements. %and 
Finds that they are a reliable predictor and can be used in real-time applications.                                                                                                                                                                                                                                                                                                                                                                                                                                                                    \\
2012                                          & Isaacs et al.                                                                                                                                                                                 &    \cite{isaacs_integrating_2012}                             & Heco                            & %This paper 
Analyzes the phenomenon called "channel blending", which is the integration of different remote and co-located interactions taking place on multiple channels at the same time. Based on an in-depth shadowing study, researchers observed that during remote interaction co-located people can also enter and leave the talk especially in close-knit groups. In such moments, the co-located participant, who is interacting with the remote one, takes the responsibility of blending the channels.                                                                                                                                                                                                                                       \\
2012                                          & Lee et al.                                                                                                                                                                                                            & \cite{lee_knowledge_2012}                                         & HT                              & %This paper 
Focuses on different types of knowledge transferred among the partially distributed conceptual design teams. %Based on an 
Experiment %, authors found 
finds that compared to other design phases, communicating during the concept generation phase is the most difficult%one
.                                                                                                                                                                                                                                                                                                                                                                                                                                                                       \\
2011                                          & O'hara et al.                                                                                                                                                                                                                       &                             \cite{ohara_blended_2011}                                         & HIM                              & %The article 
Describes the concept of "Blended Interaction Spaces" which "faithfully incorporate geometrical properties and configuration of space". %At the example of one such environment, several challenges for the creation of blended environments are described. A first set of 
Challenges include %deals with 
interaction proxemics (i.e., the link between different interaction mechanisms and how people are positioned towards other people or "information resources")%. A second set with
, and "how to map {[}the data space{]} onto the geometric properties of the space envisioned". %The authors conclude that 
Apart from %the 
technical setup, room architecture and furtniture should also be considered when tackling the two challenges.            \\
2011                                          & Pongolini et al.                                                                                                                                                                                                   & \cite{pongolini_global_2011}                                         & HM                              & %Based on 
Ethnographic research of technology experts in a multinational automotive manufacturing company, %authors 
report that a number of social and contextual factors play a major role in the choice of media, therefore showing that the task-media fit of prior theories falls short of explaining the media choices. Multiple media can be used at the same time rather than in parallel.                                                                                                                                                                                                                                                                                                                                                            \\
2011                                          & Slovák et al.                                                                                                                                                                                                     &         \cite{slovak_exploring_2011}                                                            & SI                              & Through a qualitative study of game play groups, %authors found that 
participants %had 
are found to have higher game scores %in 
when in person, but %the 
specialized group-to-group communication software was still superior to standard video conferencing tool. The authors attribute this to trust and group identity.                                                                                                                                                                                                                                                                                                                                                                                                                                                      \\
2011                                          & Yamashita et al.                                                                                                                                                                                             & \cite{yamashita_improving_2011}                                         & HIM                              & Introduces a new approach called "remote lag" to eliminate the invisibility of remote gestures, researchers found that the negative impacts of invisibility problems (of the remote collaborators) were significantly reduced.                                                                                                                                                                                                                                                                                                                                                                                                                                                                                                                         \\
2011                                          & Yamashita et al.                                                                                                                                                                                      &               \cite{yamashita_supporting_2011}                                                       & HIM                              & Analyzes multiparty fluid tabletop activities in groups of four remote people. %the study shows 
Finds a clear benefit of using upper body views. %It also shows what lead to the 
Increase in effectiveness attributed to %(e.g. 
achieving joint perspectives and leading remote collaborators to specific areas.                                                                                                                                                                                                                                                                                                                                                                                                                                                             \\
2011                                          & Yarosh \& Abowd                                                                                                                                                                                                                                      &            \cite{yarosh_mediated_2011}                                                           & HM                              & Based on interviews with 14 work-separated parents and their kids (7-13), authors detail on a number of strategies the families apply. E.g. parents use asynchronous and synchronous communication, children seek more contact to co-located adult and rely heavily on them to establish communication with the separated parent.                                                                                                                                                                                                                                                                                                                                                                                                                       \\
2010                                          & Bos et al.                                                                                                                                                                                & \cite{bos_shared_2010}                                            & SI                              & Through an experimental task called ``Shape Factory'' with U.S. college %researchers studied groups from the same 
fraternity or sorority groups, it is argued that 
shared group identity %can be very 
is beneficial for collaboration and coordination especially in extensive projects, but still cannot eliminate all the issues. Co-located and remote members diverge ``in performance, group efficacy, and sense of group identity''.                                                                                                                                                                                                                                                                                                            \\
2010                                          & Kirk et al.                                                                                                                                                                                                                                        &          \cite{kirk_home_2010}                      & HM                              & %Based on a 
Diary and interview study of home video communication users reports %researchers show 
how users achieve closeness, closer connection with family/friends, devotion and moral order at home, and list some design implications for home video communication technologies.                                                                                                                                                                                                                                                                                                                                                                                                                                                                             \\
2010                                          & Tang et al.                                                                                                                                                           & \cite{tang_threes_2010}                                         & HIM                              & Focuses on three channels of communication (person, reference, and task-space) in two studies. %authors list 
Proposes the advantages of different arrangements of users in a distributed workspace and further discover the role of ``identity, awareness, spatial metaphor, and corporeal embodiments in three-way distributed collaboration''.                                                                                                                                                                                                                                                                                                                                                                         \\
2010                                          & Turner et al.                                                                                                                                                                                                &      \cite{turner_exploring_2010}                                                                 & Heco                            & Studying the usage of communication technologies in a small company in the U.S. over a year, %researchers analyze 
finds that new communication tools do not mean leaving the old ones, and diverse tools are used for different purposes. %Their 
Communication ecosystems shape %their 
communication choices and behaviors.                                                                                                                                                                                                                                                                                                                                                                                                                                      \\
2009                                          & Ocker et al.                                                                                                                                                                                           &             \cite{ocker_training_2009}                                                         & HT                              & %Having conducted a 
Large international study reports %, authors present the 
results based on %the 
quantitative analysis %(surveys), 7 point semantic differential 
of %the 
coordinated activity variables such as ``awareness'', ``shared identity'', ``trust'', and ``team performance'' before and after training. Training %based on 462 %of the 689 participants they got results from. They 
%finds that training in teamwork %compared the results from cycles 1 \& 2 (without training) to cycle 3 (with training) and found that all variables %
significantly increasess all variables.                                                                                                                                                                                                                                                                                                  \\
2009                                          & Quinones et al.                                                                                                                                                                                & \cite{quinones_bridging_2009}                                         & SI                              & %Based on 
Reports on two case studies of undergraduate engineering students in the U.S. collaborating with students in Brazil, Israel or Turkey. %and combining surveys, interviews and observations, authors argue  that some 
Different mental models are found concerning %the 
team structure, task processes, social conventions, and about knowledge/experiences% were reported
, which led to problems. However, when the teams %managed to 
worked out the differences, harmonious collaboration %(
and even %with cases of 
friendships emerged.                                                                                                                                                                              \\
2008                                          & Tan \& Kondoz                                                                                                                                                                                                                                                                    &        \cite{tan_barriers_2008}                                                              & HIM                              & %Based on the usage of a
Reports use of ``Virtual Collaborative Desk'' by a software design team. %researchers list quantitative and qualitative findings. 
Quantitative findings were that duration of meetings shortened but meetings %got become 
became more frequent, and team participation and cultural differences emerged. %: People from Oslo had a more egalitarian participation, while Brighton was more hierarchical. 
Qualitative findings showed organizational barriers (management issues, and accessibility and policy issues) and technological barriers (hardware and software performance, network link, video, audio (less than video) and interaction devices such as pen and mouse). Suggests %Importantly they claim the 
need for real world studies rather than experiments to understand issues with virtual collaboration. 
\\
2008                                          & Yamashita et al.                                                                                                                                                                                      &                      \cite{yamashita_impact_2008}                                                & Heco                            & %This paper studies 
Reports the effects of changing seating positions of video-mediated communication (%. The authors experimented "four-person group communication: two-by-two at identical locations: 
distant parties seated across from each other vs. distant parties seated side-by-side). %They found that "in the latter
Being seated side-by-side led to more speaker switching being %were 
more evenly distributed %between distance-separated participants and co-located participants at points 
without verbal indication of the next speaker. Participants shared a higher sense of unity and reached a slightly better group solution.
\\
2007                                          & Batcheller et al.                                                                                                                                                            &          \cite{batcheller_testing_2007}                      & HIM                              & Experimented while playing the game ``Mafia'', participants in the video condition found it easier to interact with those on their side than those in the co-located condition% did
. People in the video condition found people on their side more trustworthy. Levels of fun, satisfaction and frustration remained constant between conditions.                                                                                                                                                                                                                                                                                                                                                                                                           \\
2007                                          & Chan et al.                                                                                                                                                                                                        &                    \cite{chan_facilitating_2007}                                                  & HIM                              & %Authors present their experiences in an 
International project at two universities (%located in the 
U.S. and Australia) reports %based on 
four levels of learning: personal/technical skill-based learning, topical learning, critical thinking, lessons in interpersonal communication. %They further 
Describes %some 
technical, administrative, cultural challenges%); and close with their lessons learned.
.
\\
2007                                          & Roussel \& Gueddana                                                                                                                                                                                                                                   & \cite{roussel_analog_2009}                                         & Heco                            & %The authors 
Discusses two different interpretations of ``%  the seminal 1992 paper "
Beyond being there'' by Hollan and Stornetta, concerning the term "Beyond" to mean either ``greater than'' or ``other than''. While the former interpretation suggests that by improving the fidelity of mediation solves any remaining problems, the latter states that other approaches are needed, e.g., by making possible what is impossible in-person %  face-to-face settings 
(such as additional awareness or privacy mechanisms). %Finally, they 
Proposes the concept of ``multiscale communication'' to reflect the second notion.                                                                                                                                               \\
2007                                          & Wong et al.                                                                                                                                                                                            &                             \cite{wong_sharing_2007}                                         & SI                              & %This paper is about the 
Remote expertise study %helping two workers and some of the findings show 
finds that the structure of conversation was different depending on if both workers had the same task, and %. There were also 
different types of attention requests.                                                                                                                                                                                                                                                                                                                                                                                                                                                                                                      \\
2006                                          & Bos et al.                                                                                                                                                       & \cite{bos_collocation_2006}                                         & SI                              & Through the experimental task called ``Shape Factory'', authors found that co-located doubles were victims to co-located blindness whereas remote doubles benefited. Co-located players were more embarrassed to ask for more money when they were selling to their co-located ones, but remote players didn't feel that social pressure.                                                                                                                                                                                                                                                                                                                                                                                                                      \\
2006                                          & Huang \& Ocker                                                                                                                                                                                                 & \cite{huang_preliminary_2006}                           & SI                              & Studying 12 distributed student teams, authors find that geographical distance, power and information flow play a substantial role in shaping the ``us vs. them split''. ``Work-ethic, quality of work and the mix of media in communication and collaboration'' shaped the team unity.                                                                                                                                                                                                                                                                                                                                                                                                                                                                     \\
2006                                          & O'Hara et al.                                                                                                                                                                                                                                     &            \cite{ohara_everyday_2006}                                                          & HM                              & %Based on a 
Diary and interview study %, authors argue 
finds that people make video calls for small talk, showing things or functional talks to achieve a particular goal. ``Social barriers'' to video-calling exist such as ``blurry public-private boundaries'' or not being able to show co-located ones fully, whereas examples for ``practical barriers'' can be ``noise, lighting and dual tasking''.                                                                                                                                                                                                                                                                                                                                                           \\
2005                                          & Bos et al.                                                                                                                                                                                & \cite{bos_traveling_2005}                                         & SI                              & %Researchers found that 
Changing status from remote team member to co-located team member is easier than changing from co-located to remote. The difference is due to lack of %s is easy, but the opposite is not the case because %, however for 
%co-located team members%, becoming a remote participant is not easy since they are not 
preparation for the change to remote work.                                                                                                                                                                                                                                                                                                                                                                                                                                                                                                                                                            \\
2005                                          & Mueller \& Agamanolis                                                                                                                                                                                                                                                                &                          \cite{mueller_sports_2005}                                            & Heco                            & See below.                                                                                                                                                                                                                                                                                                                                            \\
2004                                          & Bos et al.                                                                                                                                                                                       & \cite{bos_-group/out-group_2004}                                         & SI                              & %This paper is about a 
Reports on a simulation game aimed %for 
at observing collaboration in distributed teams consisting of 5 co-located and 5 remote participants. %Research has shown that 
An in-group emerged among co-located participants, who excluded the remote participants (aka ``isolates''). However, isolates also formed %an 
their own in-group.                                                                                                                                                                                                                                                                                                                                                                                                                              \\
2004                                          & Kethers et al.                                                                                                                                                                                                           & \cite{kethers_remote_2004}                                          & HM                              & %This paper is about 
Reports on a simulation %tool called 
using ``NetMeeting'' %used 
for %the 
remote meetings between researchers and %the 
farmers. %The paper 
Focuses on the asymmetries in interaction between these two groups and %authors 
found that the researchers %had 
lacked %of 
awareness of the %regarding the 
room in which the farmers were co-located, which made video more important for the researchers, while farmers did not feel the need to see the researchers since the knowledge they shared was more important than their appearance.                                                                                                                                                                                                                                                    \\
2004                                          & Powell et al.                                                                                                                                                                                                           & \cite{powell_virtual_2004}                                            & HT                              & %This is a 
Literature review %paper 
on virtual teams. %There are many interesting findings, however what we find particularly important is that the authors
Describes %mention "
the disconnect between the controlled setting and field research on virtual teams. %They underline that 
Nearly 90 percent of controlled experiments have less than eight people in team size. Also experiments allow short-term observations only, while %in 
field studies %there are 
provide more opportunities for long-term observations. % of virtual teams.
%Interestingly they do not mention a 
Hybrid settings are not mentioned in any reviewed paper.                                                                                                                                                                  \\
2003                                          & Everitt et al.                                                                                                                                                          & \cite{everitt_two_2003}                                          & HIM                              & %Authors 
Presents a remote collaboration tool called "Distributed Designers’ Outpost", supporting %activities such as distributed 
brainstorming through physical post-its and their digital representations. %Authors come up with 
Two novel awareness mechanisms are developed for distributed collaboration: "transient ink input for gestures and a vision-tracked stylized shadow for presence".                                                                                                                                                                                                                                                                                                                           \\
2003                                          & Hutchinson et al.                                            &         \cite{hutchinson_technology_2003}                                                             & Heco                            & %Based on a 
Cross-cultural study on families (%in 
France, Sweden and the U.S.) that %, authors 
evaluates a new method %they developed 
for co-designing technology with users, called "technology probes". %They analyze 
Two technology probes are reported, the messageProbe and the videoProbe. Results show that %they are successful in three ways: 
technology probes bring to light practical needs and fun among families, they relied on real-life scenarios, and boosted creativity by introducing new technologies.                                                                                                                                                                                                                                                    \\
2003                                          & Mueller et al.                                                                                                                                                                                               &           \cite{mueller_exertion_2003}                                                           & Heco                            & %Researchers 
Develops an Exertion Interface using a life-size video-communication system, enabling two remote people play a soccer ball and compared the same experience with a non-exertion keyboard interface. %Based on a questionnaire and interviews with 56 participants, authors argue that in general 
EI users got to know each other better, became closer friends, and enjoyed the game more compared to the non-EI users. %Apart from the study's findings, authors mention some 
Provides guidelines for future Extertion Interfaces.                                                                                                                                                                                                                       \\
2001                                          & Geyer et al.                                                                                                                                                        & \cite{geyer_team_2001}                                            & Heco                            & %Authors 
Describes %their 
collaborative workspace prototype, called Teamspace, which supports both synchronous and asynchronous team collaboration. Describes %They also mention its 
design issues based on their experiences, %with the tool. They 
and presents issues around security, meeting capture, and awareness.                                                                                                                                                                                                                                                                                                                                                                                                                                                 \\
1998                                          & Mark                                                                                                                                                                                                                                                   & \cite{mark_building_1998}                                          & SI                              &  Building trust and relationships in virtual teams is improved when there are facilitators for social and technical support, chat for informal communication including private jokes, flexibility of location, data sharing from personal computers, and %face to face 
in-person contact and meetings.                                                                                                                                                                                                                                                                                                                                                                                                                                                       \\
1998                                          & Monk \& Watts                                                                                                                                                                                                                                              &                      \cite{monk_peripheral_1998}                                                & HM                              & Remoteness does not play a role in the sense of social presence, but being peripheral does. %Also they found out that in 
When recalling a conversation or activity, being peripheral is not decisive. Sometimes people can be peripheral and do not actively participate but still are interested in the activity or discussion. Interest is the key factor.                                                                                                                                                                                                                                                                                                                                                                                               \\
1998                                          & Morikawa \& Maesako                                                                                                                                                                                                                                   & \cite{morikawa_hypermirror:_1998}                                         & Heco                            & Second study on HyperMirror. While using the tool, participants mostly referred to their presence in the virtual space rather than their presence in the physical room. %Authors further analyze 
Different aspects of the sense of reality include gaze direction, feeling of personal space, and approaching the camera.                                                                                                                                                                                                                                                                              \\
1997                                          & Inoue et al.                                                                                                                                                                                                                  &       \cite{inoue_integration_1997}                                                               & Heco                            & Reports on a %Authors designed and tested a 
video communication tool called HERMES, which allows each co-located participant sitting around a round table, seeing remote participants in front of their seat through a monitor. %Authors studied 
Speech and gaze directions of the participants %and the results 
were in line with the aim of the design, showing the importance of seat and monitor arrangement in video-based communication. However, speaker identification through loudness of the voices was not realistic.                                                                                                                                                                                                                                          \\
1997                                          & Morikawa \& Maesako                                                                                                                                                                                                                                              &              \cite{morikawa_hypermirror:_1997}                   & Heco                            & %Through a WYSIWIS system called 
Reports on the `HyperMirror' system that provides the feeling of ``being in the same room''. System is symmetrical and easy to understand, minimizing psychological barriers between participants. %, which provides "being in the same room" feeling, authors mention when 
Finds that the sense of reality can increase or decrease (e.g., when participants walked into each other on the mirror). Lack of %They also claim that there was not 
haptic feedback was noticeable, because the remote persons were only virtually connected.                                                                                                                                                                                                                                                                                                                                                                                                             \\
1994                                          & Kuzuoka et al.                                                                                                                                                                                                        & \cite{kuzuoka_gesturecam:_1994}                                         & HIM                              & Bad quality image and video transmission delay create problems in instruction contexts. Wider screen can be useful for %the 
instructors based on the task. Language, humor, portability and control by a master actuator helps users feel sympathy towards the system.                                                                                                                                                                                                                                                                                                                                                                                                                                                                                                             \\
1994                                          & Oliver                                                                                                                                                                                                                                                      & \cite{oliver_software_1994}                                         & HIM                              & CMC improves the sense of community among distance learners, however, lack of equity in varying sized groups, non-effective group members and technical difficulties cause problems.                                                                                                                                                                                                                                                                                                                                                                                                                                                                                                                                                                    \\
1988                                          & Ahuja et al.                                                                                                                                                                                                                                  &         \cite{ahuja_rapport_1988}                                                             & Heco                            & More flexible input controls, larger storage of information and better devices are needed. Long-distance conferencing is still challenging.                                                                                                                                                                                                                                                                                                                                                                                                                                                                                                                                                                                                            
% \end{tabular}%
% }
\end{longtable}

%\clearpage
\vspace{1.5cm}
{\Large \textit{\textbf{Part B - Additional Findings and Synthesis}}}
\label{sec:partB}

\section{What's Missing}
\label{sec:missing}
%\todo{TN: We could create this section with a two-fold function: to firstly describe omissions similar to a limitations section and secondly present important work future researchers should not be missing in addition to the systematic status quo described above.}

SLRs have some inherent limitations which lead to missing relevant publications, such as focusing on specific online libraries/catalogs, using \textit{a priori} keyword search queries which potentially lead to keyword mismatches with relevant literature, and having to choose keyword terms which are broad enough to capture the relevant literature while also limiting false positives. These limitations are exacerbated by terminological confusion as pronounced as it is this case. Part A of this review provided descriptions of what could be systematically found in the ACM DL using the terms most likely to be of relevance. In Part B, we extend these results with other related work identified in our own prior research, and, in discussion of key themes, synthesize this work with the SLR findings. First, we describe two sets of notable exclusions from our final SLR (i.e., they were not part of our initial set of 1,209 results, due to the reasons stated above) that are essential for a full picture of prior HCM research.

%Through this combination we aim to provide an extensive resource for shaping the future of HCM development. Moreover, we present these additional works in our taxonomy to facilitate the identification of particular aspects designers of future HCM need to conceptually include in their considerations.
%\subsection{Notable Exclusions}
%In order to minimize the possibility of missing major relevant publications, which fit our inclusion criteria but still do not show up in our set of results, we consulted several %top experts
%\tn{senior researchers} specialized in this field and received written feedback from them including some paper suggestions as well.

%either excluded (i.e., they were in our initial set of 1,209 results but we actively decided for exclusion) %\todo{We do not have exclusions until now, only papers that were not included, so maybe change this when we know about all papers} or 
  %For an overview of these cases, please refer to Table \ref{tab:notable_exclusions}.

%\textcolor{red}{Should we actively tell our readers that several top experts in the field judged our inclusions (after all, this could be regarded as some kind of written expert interview or expert consultation)? And if so, to which degree should we specify who they were?}
%Some of the publications in this list were recommended by top authors/a top author in the field of HCM (see Section \ref{sec:authors}), others were known among the authors of this article from previous narrative literature reviews.

\subsection{Publications on Media Spaces}

Perhaps the most notable exclusion in this SLR is most of the research exploring Media Spaces (except \cite{ohara_blended_2011}). This research exists in articles in the ACM DL, but also in some non-ACM outlets, and in edited collections~\cite{finn_video-mediated_1997, harrison2009, neustaedter_connecting_2013}\footnote{The edited collections are listed in the ACM Full-Text Collection but not the individual chapters, although some chapters are direct copies or revised versions of conference papers or journal articles}. Media Space research arose from the recognition that cooperative work involves awareness of integrated joint context that seamlessly flows between individual and group, between focus and spontaneity, between productivity and sociality, all contained within \textit{architectural spaces}~\cite{kraut_patterns_1988, heath_collaborative_1991, schmidt_cooperative_2011}.

The term `Media Space' was coined by Robert Stults~\cite{stults_mediaspace_1986} as a ``computing/video setting for unstructured collaborative work among people separated by space and time.'' Stults never uses the term `hybrid', nor does it appear in most of the subsequent Media Space literature\footnote{Indeed, `hybrid' does not appear in the abstracts or keywords of Media Space articles (as noted above), nor the indexes of three major edited collections entirely or partially exploring Media Spaces~\cite{harrison2009, neustaedter_connecting_2013, kraut_intellectual_2014}. The one appearance of `hybrid' in an edited collection index (in~\cite{finn_video-mediated_1997}) refers only to communication network topology~\cite{hiroshi_ishii_iterative_1997}. }. This omission can be ascribed to the recent rise of the term's popularity, but also because, conceptually, Media Space settings were intended to \textit{naturally and seamlessly include hybrid cohorts}, so there was no need to describe them as such. Stults clearly intended to cover `hybrid' collaboration, e.g., the photograph on the first page of his report~\cite{stults_mediaspace_1986} depicts a group of four people working together, two in a local room and two working remotely and connected via individual monitors on the local desk. As a term and a research concept, `Media Space' largely fell out of favour by 2015, but it has been kept afloat by Steve Harrison, Carman Neustaedter, and colleagues/students exploring their use in domestic contexts. As of October 2021 in the ACM DL, the latest paper to use `Media Space' in the title is a CHI2020 poster~\cite{Cedeno-Mieles_etal_2020}, and before that, a CSCW 2014 Companion paper~\cite{odour_etal_2104}. Neustaetder et al.~\cite{neustaedter_sharing_2015} summarizes much of this research without using `Media Space' in the title. However, interest is returning to the concept in the wake of \tn{global reactions to} COVID-19 \tn{and resulting} harsh lessons on disconnection from the fabric of collegiality~\cite{weiley_losing_2020}.

\sr{A significant amount of Media Space research sought to enable \textit{both} fully remote and hybrid conditions. For example, Buxton's "Meetingspace -- Mediaspace -- Meaningspace"~\cite{buxton_mediaspace_2009} proposes that three spaces for the mechanics of communication (person, task, and reference spaces) need to be supported for any form of geodistribution. In such a case, supporting hybrid geodistribution usually also solves for fully-remote geodistribution, at least in terms of the embodied mechanics of meaning. Similar research from edited collections includes both the organizational context (or more-or-less agnostic) \cite{luff_creating_2009, churchill_media_2009, karahalios_social_2009, henderson_videoconferencing_2009, gorzynski_halo_2009, ensor_virtual_1997, buxton_living_1997, buxton_etal_interfaces_1997, hiroshi_ishii_iterative_1997, wilbur_models_1997} and the domestic context \cite{inkpen_kids_2013,judge_inter-family_2013, neustaedter_sharing_2015}.} 

\sr{So, an undercurrent of hybridity runs through Media Space research, but both terminological and conceptual difficulties can make it difficult to rule them in or out \textit{of an SLR}. At a base level, searching for the phrase produces huge numbers of false positives because either the words may appear adjacently with or without compound meaning. But even within the identifiable `Media Space' literature there are problems. For example, Schmidt~\cite{schmidt_problem_2002} notes that despite `awareness' being a key goal of much Media Space research, the definition is not consistent. The relationship of Media Space awareness to other `awarenesses' is not clearly specified, and `awareness' is also often prefixed with adjectives to produce more nuanced terms, such as `collaboration awareness', `peripheral awareness', `background awareness', `passive awareness', `reciprocal awareness', `mutual awareness', and `workspace awareness'. These different awarenesses are differently relevant to HCM, and thus finding articles in which these terms are used in the context of hybridity is challenging if the goal is to understand what is special about hybridity.}

\sr{Some seminal Media Space articles provide critical nuances to hybridity. For example, Harrison \& Dourish's \cite{harrison_re-place-ing_1996} ``Re-Placing Space'' explores the conceptual links between space and place in the lived reality of work, and how the Media Spaces that succeed move beyond pure connection between spaces and instead enable some of the crucial elements of what local groups in hybrid situations build together, such as the artifacts of shared history. However, they also point out some 
`placeless spaces', such as USENET newsgroups, can build shared histories with purely digital artefacts with no reference to real spaces or, indeed, hybrid groupings. While Media Spaces may be seen as the obvious panacea to the lack of geodistributed collegiality, they also either introduce serious challenges, especially to privacy~\cite{boyle_privacy_2009, friedman_watcher_2009}. Media Spaces -- especially those that involve hardware -- can become highly complex as personal and shared ecosystems of devices increase in size and complexity~\cite{roussel_analog_2009}, leading to adoption problems if the systems are also not flexibly fitted to practical issues of daily use~\cite{aoki_section_2009, roussel_analog_2009, karahalios_social_2009, bly_disconnecting_2009, alem_presence_2009, corrie_build_2009, kristensen_media_2009}. This is especially evident in the context of increased mobile device use~\cite{aoki_bringing_2009, ohara_media_2009}.}

\sr{In ``Media Space, After 20 years'', Stults~\cite{stults_media_2009} reflects that the initial imperative for the 1980s `Media Space' concept was that Design as a professional activity relied on the materiality of the physical space of design studios and physical artifacts of both tools and outcomes. However, by 2008, he acknowledged that the digital realm had increased in scope, leading him to question the primacy of physicality. In a similar spirit, Greenberg et al.~\cite{greenberg_reflecting_2009} reflect that the metaphors underpinning the design of many Media Spaces tended towards the static rather than the mobile, and, ironically, as digital connection technologies improved, also tended to silo digital and non-digital interaction rather than blend them or enable smooth transitions between them. This is especially the case for what Stults (indirectly) and Greenberg et al. (directly) posit as a \textit{unique value} of hybrid Media Spaces, which is the affordance of informal opportunistic engagement. In the post-COVID19 era, the need to enable both fully-remote and hybrid informal collegiality throughout a day has become imperative, which speaks to an acknowledged need for the affordances of physicality. One tranch of Media Space research points to how this might be enabled, and we will discuss this below in the Focal Points (Section \ref{sec:focal}).}

\subsection{Publications on Partially Distributed Teams}
The research on partially distributed teams (PDTs) has been identified previously as a valuable source of information for HCM. Neumayr et al. \cite{neumayr2018} list five publications on this area and provide a brief overview of their findings. Three of these (\cite{bos_traveling_2005,bos_collocation_2006,bos_shared_2010}) are part of our SLR. Additionally, four PDT papers which were not part of the compilation in \cite{neumayr2018} resulted from our \tn{systematic approach in part A} and have consequently been included \cite{bos_-group/out-group_2004,huang_preliminary_2006,lee_knowledge_2012,lee_input-process-output_2013}. 

However, there are still other PDT publications that were not included in our SLR. One example is ``Cross-cutting faultlines of location and shared identity in the intergroup cooperation of partially distributed groups'', which  describes several challenges of partially distributed groups especially relevant to HCM such as the interplay between having a shared identity and one's location \cite{voida2012}. The paper was not included/selected in our initial set due to keyword mismatches with our search query. The authors used `partially distributed groups' or `partially distributed work' to describe their subject matter. While `partially distributed' is part of our search query's first set, it was only triggered if any of the query's second set's terms was used in combination, which was not the case. %Furthermore, in the abstract, they used the rather general terms `distributed collaboration' to refer to the remote collaboration between two subgroups in their study. 
Another example is Cheshin et al. \cite{cheshin2013}, who investigated for PDTs how electronic communication norms (ECNs) emerge and furthermore ``showed that traveling members kept some of their ECNs after swapping remotes and collocateds'' \cite{neumayr2018}. The article was not included because it was published in the Journal of Personnel Psychology (Hogregfe), outside the ACM DL. Overall, we have the impression that the highly related subtopic of PDTs is reasonably well covered in our final set of publications, but may need further investigation.

%\sr{NOTE TO \tn{THOMAS} - THESE TABLES ARE NOW OUT OF DATE. SHOULD WE REDO THEM?}

%\begin{table}
%\centering
%\caption{Examples of notables publications not included in the SLR.}
%\begin{tabular}{llll}
%\textbf{Reference}  & \textbf{Year} & \textbf{Not included or excluded} & \textbf{Reason} \\
%\hline
%Isaacs et al. \cite{isaacs1996} & 1996 & not included & keyword mismatch \\
%Voida et al. \cite{voida2012} & 2012 & not included & keyword mismatch \\
%Cheshin et al. \cite{cheshin2013} & 2013 & not included & not in ACM DL \\
%Karis et al. \cite{karis2016} & 2016 & not included & not in ACM DL \\

%\hline
%\end{tabular}
%\label{tab:notable_exclusions}
%\end{table}

%\begin{figure}[ht!]
%  \includegraphics[width=1\linewidth]{figures/Taxonomy_Non-Inclusions.pdf}
%  \caption{Example of how the non-included publications would fit into the taxonomy.}
%  \label{fig:taxonomy_non_inclusions}
%\end{figure}

%\sr{NOTE: I've rolled up the 'Seminal work' section into the notable exclusions/non-inclusions' section}

%\subsection{Further Seminal Work}
%\tn{In this section, we give an overview of the most important resources researchers and designers of future HCM systems should definitely consider in addition to the results of the systematic approach.}
%\todo{Books, papers, articles based on the list to be compiled by Sean.}

%\paragraph{\todo{Influential Book/Paper II}}

%\paragraph{\todo{Influential Book/Paper III}}

% \renewcommand{\arraystretch}{1.7}

% \renewcommand{\arraystretch}{1}

\section{Focal Points}
\label{sec:focal}
Based on the SLR, we propose the following focal points for future investigation. We synthesize literature both from the systematic approach (summarized above in \autoref{tab:findings}) and additional literature (from \autoref{sec:missing}) as a summary of what we already know and pointers towards open questions.

\subsection{Hybrid Interaction Mechanics}
Our first focal point on hybrid interaction mechanics deals with the aspects that make interacting in hybrid settings special. These aspects often arise from hybrid collaboration situations because of their \textit{mix of synchronous and asynchronous work, because they span multiple groupware applications and devices, and because some people are in the same place and some remote} \cite{neumayr2018}. This definition (see also \autoref{sec:def_hybrid_collaboration}) also inspired us to shape the taxonomy presented in \autoref{sec:taxonomy}. There are multiple points of interest in investigating hybrid interaction mechanics, such as how people sharing an office space can be aware of each other's activities and more easily ask quick questions of each other. Some of these issues are covered in the Media Space research, but there is more to be understood about the special characteristics of collaborative coupling and transitions during mixed-focus collaboration.

\subsubsection{SLR}
Some pointers in our review can be found in O'Hara et al.'s article ``Blended Interaction Spaces for Distributed Team Collaboration'' \cite{ohara_blended_2011} where the concept of media spaces is extended to blended interaction spaces, and provides several examples of work patterns of mostly synchronous interactions in hybrid settings. The authors focus on the embodiment of interactions in space and how to represent their spatial geometries consistently over distance. Neumayr et al.'s Domino Framework \cite{neumayr2018} present a way of describing and analyzing a multitude of different hybrid settings. They describe the phenomenon of team fragmentation, which results from teams frequently splitting up into subgroups (see also our discussion of behavior towards forming subgroups in \autoref{sec:taxonomy_subgroups}). 

Papers related to hybrid interaction mechanics included in our SLR can be found here (in chronologically ascending order): \cite{oliver_software_1994}, \cite{kuzuoka_gesturecam:_1994},  \cite{everitt_two_2003}, \cite{chan_facilitating_2007}, \cite{batcheller_testing_2007}, \cite{tan_barriers_2008}, \cite{tang_threes_2010}, \cite{yamashita_improving_2011}, \cite{yamashita_supporting_2011}, \cite{ohara_blended_2011}, \cite{damian_instructional_2012}, \cite{salimian_exploring_2015}, \cite{licoppe_showing_2017}, \cite{neumayr2018}, \cite{fu_teaching_2018}. A summary of their findings is presented in \autoref{tab:findings} in lines where the Focal Point (FP) column tag is ``HIM''.

\subsubsection{Additional References}
Another lens to view collaborative interaction is through the mechanics of shared understanding, such as the task, person, and reference spaces suggested by Buxton \cite{buxton_mediaspace_2009}. This kind of stance is valuable for analyzing hybrid interaction mechanics in comparison and contrast with findings of more co-located (see e.g., \cite{long_2017,krogh_2017}) or remote collaboration (see e.g., \cite{junuzovic_2012,pejsa_2016,piumsomboon_2018}). 

\subsubsection{Future Relevance}
Although many prototypes have been proposed to help with establishing a shared reference space among local and remote participants (e.g., \cite{gronbaek2021}), we are not there yet in commercial systems. Similarly, bringing analog artifacts (which are usually accessible only to local participants) to the attention of remotes is still cumbersome with current widespread tools, as Licoppe et al. showed \cite{licoppe_showing_2017}. When designing for future hybridity, we should expect teams splitting up into subgroups, which may enable division of labor but can be disadvantageous when a task requires a closer coupled collaboration among larger groups. We feel that the spontaneous breakout of subgroups in larger meetings is still not supported well enough in remote and even more in hybrid settings, when subgroups would like to form with several co-located and remote persons. 

%Another interesting work by O'Hara et al. \cite{ohara_blended_2011} focuses on distributed set-ups blending the spatial geometries of two remotely connected rooms to provide a unified spatial frame of reference, which can increase the embodiment in video-based communication among hybrid team members. 

\subsection{Hybrid Meetings}
Hybrid meetings refer to video- or audio-based meetings involving both in-room and remote attendees \cite{saatcci2019hybrid}. As described in Section \ref{sec:def_hybrid_meetings} more in detail, there has been a lack of unity in term usage describing this format% and terms such as distributed or partially distributed meetings have been used in the HCI and CSCW community widely
. 

\subsubsection{SLR}

%Therefore, we mentioned that before 2020 
The only paper included in our SLR referring to this meeting format as hybrid meetings is the paper from 2017 by Xu et al. ``Attention from Afar: Simulating the Gazes of Remote Participants in Hybrid Meetings'' \cite{xu_attention_2017}. This paper focuses on improving social cues at hybrid meetings by simulating the gazes of remote participants for local ones \cite{xu_attention_2017}. Authors find that using simulated gaze in hybrid meetings has its pros and cons. While the feeling of presence was increased, simulated gazes do not always correctly reflect what the remote participants pay attention to \cite{xu_attention_2017}. 

Other papers related to hybrid meetings and included in our SLR are listed here: \cite{monk_peripheral_1998}, \cite{kethers_remote_2004}, \cite{ohara_everyday_2006}, \cite{kirk_home_2010}, \cite{yarosh_mediated_2011}, \cite{pongolini_global_2011}, \cite{hradis_voice_2012}, \cite{barden_telematic_2012}, \cite{yarosh_almost_2013}, \cite{neustaedter_sharing_2015}, \cite{awori_sessions_2016}. A summary of the findings can be found in \autoref{tab:findings} in lines where the Focal Point (FP) column tag is ``HM''.

\subsubsection{Additional References}

Le et al.'s ``DigiMetaplan: Supporting Facilitated Brainstorming for Distributed Business Teams'' uses the term `hybrid' when referring to ``hybrid teams'' \cite{le2019}. This paper is highly relevant, and is missing only because it was published after our SLR query of the ACM DL. %\tn{(please note that there is another paper \cite{le2019} using hybrid in this way but not reporting about ``hybrid meetings'' but ``hybrid teams'' instead; the paper---although highly related---was however published only briefly after we queried the ACM DL)}. 
%There are other papers in this field using the term hybrid meetings and researching this meeting format explicitly such as 
Saatçi et al.'s two ethnographic papers are also highly relevant but not included due to the SLR protocol. Their most recent paper on reconfiguring hybrid meetings in the business setting \cite{saatcci2020re} was published after the SLR query, and their earlier paper \cite{saatcci2019hybrid} was not included because the ACM DL does not index the CollabTech conference.

\subsubsection{Future Relevance}

%Recently, after the COVID-19 pandemic entered our lives, hybrid meetings as a meeting format has been receiving attention more than ever and getting more and more widespread in different sectors and workplaces, leading to more diverse meeting experiences. 
We expect a rise in the amount of both experimental and real-world studies on hybrid meetings in HCI and CSCW as well as in other fields. As our SLR shows, research on hybrid meetings in the pre-COVID-19 era are conducted mostly in the laboratory setting. We believe that in the post-COVID-19 world there is a huge need for real-world studies since there are more opportunities to observe hybrid meetings in diverse workplaces. Furthermore, due to an immediate and world-wide adoption of technical skills especially by knowledge workers, we think that the experiences of participants have been evolving widely. For instance, participants of hybrid meetings got used to using videoconferencing technologies more than ever and adopted skills supporting better meeting experiences such as improving their digital literacy, getting used to virtual turn-taking, using headphones, (un)muting microphones etc. Such a wide and increasing usage of videoconferencing technologies also motivated companies developing these tools to compete with each other in testing and launching new features. All these factors show us the need for observing these emerging ``hybrid workplaces''%, the newly developed and used concept for workplaces, which have shifted to hybrid format of working and adapted their technical and social infrastructure to support HCM
. We recommend researchers to further explore the new multiplicity of hybrid meeting experiences and focus on more disadvantaged or problematic workplaces such as educational institutions, which are widely exposed to digital divide as well as involve pedagogical concerns.  %With the hope that it gives some inspiration for future research, we would like to mention some of the relevant findings from our SLR papers on hybrid meetings across different domains.

%Research on the transmission of indigenous knowledge to young population in Kenya through video-mediated communication shows that elderly people require hands-free and easy-to-use technologies while teaching remotely \cite{awori_sessions_2016}. Another research on the domestic use of videoconferencing technologies underlines the differences between workplace and home in hybrid meetings and finds that there is a strong need for moving beyond conversations in supporting long-term video-based connections among family members or close friends, who aim to share their everyday life experiences with each other as well.

%Mini definition: \textit{Mix of in-room and remote people}.  Attention From Afar, Xu et al.: \cite{xu_attention_2017}

\subsection{Hybrid Ecosystems}
Our third focal point originates from the aspect of Neumayr et al.'s definition of hybrid collaboration \cite{neumayr2018}, that hybrid teams often use various tools and devices. These hybrid ecosystems are often interwoven with a mix of devices and software both for a single individual (their home and office set-up) and for teams working together. 

\subsubsection{SLR}
Papers related to hybrid ecosystems and included in our SLR can be found here: \cite{ahuja_rapport_1988}, \cite{morikawa_hypermirror:_1997}, \cite{inoue_integration_1997}, \cite{morikawa_hypermirror:_1998}, \cite{geyer_team_2001}, \cite{mueller_exertion_2003}, \cite{hutchinson_technology_2003}, \cite{mueller_sports_2005}, \cite{roussel_beyond_2007}, \cite{yamashita_impact_2008}, \cite{turner_exploring_2010}, \cite{isaacs_integrating_2012}, \cite{garbay_normative_2012}, \cite{falelakis_automatic_2012}. For their findings, see \autoref{tab:findings} where the Focal Point (FP) column tag is ``Heco''.

\subsubsection{Additional References}
Brudy et al. \cite{brudy2019} present a survey paper of cross-device interaction, which is a comprehensive guide to learning the different facets that arise when people interact using multiple devices. Most of our synthesized papers also include at least two different software tools or hardware devices (see the Taxonomy's according columns in \autoref{fig:taxonomy} and \autoref{fig:taxonomy2}). 

\subsubsection{Future Relevance}
Important questions in this regard are how to create coherent experiences over this dispersed ecology of devices and software and how technological support can augment and help overcome typical obstacles involved in HCM (such as a lack of awareness, or strongly fluctuating modes of presence). Conceptually, we feel that general models and frameworks can guide the way to understanding these questions. For example, Buxton's media spaces \cite{buxton_mediaspace_2009} can direct us focusing on the people who collaborate (be it collocateds or remotes) and assess their individual capacity to utilize the task, person, and reference space. Many of the known problems in HCM stem from the fact that any of these spaces cannot be sufficiently utilized and this is where we think that support should be provided. 

One example for first steps in providing specific support for these spaces is MirrorBlender \cite{gronbaek2021} which provides a malleable videoconferencing system for hybrid meetings. Besides more standard videoconferencing features for the person and task space, the authors suggest WYSIWIS so that the ``the canvas looks exactly the same across diferent computer displays'' with the aim of synchronizing and blending the different camera perspectives together. In this way, it is possible to point at specific elements on a shared screen using one's own camera image for deictic gestures, effectively supporting the reference space. Although the problem of parallax still is an issue (resulting from the distance between the camera and the area on a UI one would like to point at), we think that approaches similar to this are promising to facilitate a coherent collaborative interaction among hybrid groups spanning several tools and devices in a hybrid ecosystem.

% \subsection{Hybrid Schedules}
% \todo{THOMAS integrate into Teams}
% Mini definition: \textit{people working from home may have very different schedules from those in the office}. %Temporally Distributed Teams, Barksdale et al.: \cite{barksdale_video_2012}

% A paper related to hybrid schedules and included in our SLR is Barksdale et al.'s paper on temporally distributed teams \cite{barksdale_video_2012}.

\subsection{Hybrid Teams}
%Mini definition: \textit{Used very generally but can reflect distributed work groups and what flows from that such as being in different time zones}. Bridging the Gap, Quinones et al.: \cite{quinones_bridging_2009}
%Blended Interaction Spaces, O'Hara et al.: \cite{ohara_blended_2011}

Hybrid teams refer to the distributed work groups mostly living in different time zones. The difference \tn{between} hybrid teams \tn{and} virtual teams is that there is a co-located group of people working from the same office(s) and there are remote team members, who collaborate with the co-located group. \tn{Furthermore, when different time zones do not play a role in a hybrid team, still people working from home may have very different schedules from those in the office.} 

\subsubsection{SLR}
Publications with a focus on Hybrid Teams in our SLR are: \cite{powell_virtual_2004}, \cite{ocker_training_2009}, \cite{lee_knowledge_2012},\cite{barksdale_video_2012}, \cite{siitonen_i_2013}, \cite{lee_input-process-output_2013}, \cite{bendix_role_2013}. A summary of their findings can be seen in \autoref{tab:findings} in lines where the Focal Point (FP) column tag is ``HT''.

Additionally, Quinones et al.'s paper \cite{quinones_bridging_2009}, which was also included in our SLR, is an example of research on hybrid teams in the educational domain. The authors analyzed the role of cultural differences and mental models in international teamwork by conducting two case studies on undergraduate students studying civil engineering in the United States, who collaborated with students from Brazil, Israel, and Turkey remotely. Their study brings into light the different mental models concerning the team structure, task processes, social conventions and knowledge/experiences in an international hybrid team. One paper related to hybrid schedules and included in our SLR is Barksdale et al.'s paper on temporally distributed teams \cite{barksdale_video_2012}.

\subsubsection{Additional References}
Le et al's paper ``DigiMetaplan'' \cite{le2019} besides investigating hybrid meetings also covers some aspects related to hybrid teams as they suggested digital, facilitated brainstorming for distributed teams in business contexts. 

\subsubsection{Future Relevance}
Hybrid teams have different working dynamics compared to fully co-located or virtual teams, since the differences in time, space and culture as well as the asymmetries in interaction due to the video-mediated communication may create social barriers among co-located and remote participants. We believe that after the pandemic, more research on the characteristics of hybrid teams will be necessary.

\subsection{Hybrid Patterns of Social Interaction}
%Mini definition: \textit{People in the office at the watercooler, and at home missing out from serendipitous connections and the life of the office.}
Hybrid patterns of social interaction refer to the social interactions beyond the work-related tasks among workers at a hybrid work setting. Small conversations among co-located employees while waiting in front of the coffee machine or remote workers working from home and missing out serendipitous connections and life at the office can be examples of the diversity of hybrid patterns of social interaction for co-located and remote participants. 
\subsubsection{SLR}
Papers focusing on hybrid patterns of social interaction are listed here: \cite{mark_building_1998}, \cite{bos_-group/out-group_2004}, \cite{bos_traveling_2005}, \cite{huang_preliminary_2006}, \cite{bos_collocation_2006}, \cite{wong_sharing_2007}, \cite{quinones_bridging_2009}, \cite{bos_shared_2010}, \cite{slovak_exploring_2011}, \cite{kauffmann_knowledge_2018}. \autoref{tab:findings} gives an overview of their findings where the Focal Point (FP) column tag is ``SI''. 

%Traveling Blues, Bos et al.: \cite{bos_traveling_2005}
%Collocation Blindness, Bos et al.: \cite{bos_collocation_2006}
%In-/Out-Group Effects, Bos et al.: \cite{bos_-group/out-group_2004}
%Shared Identity, Bos et al.: \cite{bos_shared_2010}
%Electronic Communication Norms, Cheshin et al.: \cite{cheshin2013}
%Prelim. In-/Out-Group Effects, Huang and Ocker: \cite{huang_preliminary_2006}
%Knowledge Sharing, Kauffmann and Carmi: \cite{kauffmann_knowledge_2018}

\subsubsection{Additional References}
We mentioned above that the `Media Space' research, which was left out of the formal SLR, nevertheless has a crucial role to play in understanding hybrid collaboration and meetings. This is especially the case in terms of hybrid patterns of social interaction that are opportunistic and often also informal. Opportunistic (spontaneous and serendipitous) and informal interactions throughout the work-day play a major role in coordination, productivity, and the well-being of groups~\cite{aiken_inf_organic_1971, marrett_inf_communication_1975}. Opportunistic talk occurs at lunch~\cite{austin_lunch_2018}, corridors/hallways~\cite{long_corridor_2007}, and by watercoolers~\cite{lin_watercooler_2015}, and it takes the form of productivity-oriented talk across desks~\cite{babapour_appropriation_2018}, over-the-shoulder-learning~\cite{twidale_OTSL_2005}, pre-meeting chat~\cite{yoerger_etl_allen_chitchat_2015}, and gossip~\cite{noon_gossip_1993}. Even when informal, such `small talk' is neither just idle or uninformative, it is the resource for actively doing collegiality~\cite{holmes_doingcollegiality_2000}. 

In hybrid work contexts, often both the \textit{pre-conditions} for being aware of others enough to have spontaneous and serendipitous engagements are missing, as are \textit{designs for comfortable encounters} that do not feel the same as scheduled or formal collaboration and meetings~\cite{aoki_section_2009}. Media Space research has explored how to deal with both issues~\cite{fish_videowindow_1990, gaver_realizing_1992, dourish_portholes_1992, fish_videowindow2_1993, isaacs_etal_informal_1997, ackerman_hanging_1997, kim_magic_2007, aoki_bringing_2009, karahalios_social_2009, greenberg_reflecting_2009}. Of course the same problems apply to fully-remote contexts, but there is a distinct difference between what each must enable~\cite{isaacs1996, tang_supporting_1994}. This is one case in which the design for hybrid does not necessarily encompass the design for fully-remote, as the exigencies of informal awareness in hybrid situations are one instance in which asymmetry is especially pronounced.

This is seen in the result of one of the earliest examples of a system that enabled informal hybrid encounters: VideoWindow by Fish et al.~\cite{fish_videowindow_1990, fish_videowindow2_1993}. Fish et al. argue that there are certain characteristics of physical proximity that a Media Space for informal communication requires: a concentration of suitable partners, co-presence, low personal cost, and a visual channel that enables long-range identification as well as an audio channel for interaction. VideoWindow used an always-on AV wall to connect the common areas of two floors of a building. To encourage a cohort of users, mailboxes were moved to the vicinity of the area and free coffee was on offer nearby. Beyond some technical issues, the largest problem turned out to be privacy: the system made people uncomfortable having private conversations in the common areas -- both those who were in-person and those using the system. Just over a decade later the Magic Window system faced similar problems~\cite{kim_magic_2007}. The system also required people to go to an area to socialize, rather than be where they were. Karahalios~\cite{karahalios_social_2009} notes that a central property of a socially-oriented Media Space is that it contains a ``social catalyst'' -- an endogenous reason for people to engage. Without a social catalyst, a Media Space for informal hybrid encounters may need to be linked more to office `neighbourhoods' than to break rooms. Karis et al.~\cite{karis2016} describe creating such `office neighbourhood' experiences by using Google's video conferencing system to create `Video Portals'.\footnote{This article was not included in the SLR because it was published in the journal Human-Computer Interaction by the Taylor and Francis Group.}

\subsubsection{Future Relevance}

Enabling opportunistic and informal encounters will be essential to enabling the success of hybrid work. The Media Space research points the way to the opportunities and challenges. A market is certainly emerging in this area, with two start-ups vying for attention in 2021: Video Window\footnote{\url{https://videowindow.com/}, last access October 27th, 2021% (amusingly) bills itself as "The world’s first always-on immersive video conferencing portal".
} and Perch\footnote{\url{https://perch.co/}, last access October 27th, 2021}. Furthermore, as the hybrid workplace has been envisioned for the post-COVID world, and many companies or educational institutions have already started or are planning how to adapt to this setting, there will be many opportunities to further observe and research the social interactions among colleagues in a hybrid work setting.

\subsection{Hybrid Events}
%\todo{Special kind of context, self-containedness, Larger meetings, happening once, more crowded, maybe different hybridity such as all co-located but can at the same time use devices to connect to others. Or is this only one application scenario?}
%\todo{Hybrid Presentations (as in lecture or product presentation)}

Before the COVID-19 pandemic, hybrid events could be considered as co-located events such as conferences, conventions, fairs or seminars, which also make use of digital tools and \emph{backchannels} to support and augment the interaction among the co-located participants \cite{nelimarkka2018hybrid}. However, recently after the COVID-19 pandemic entered our lives, hybrid events as a term has been also widely used for referring to large events involving both co-located and remote participants. The conference tourism industry is contemplating how to make conferences, conventions, and fairs sustainable in the shadow of COVID-19. Even though hybrid events may involve hybrid meetings, they are clearly a separate category  ---  larger, spread out over time, and often cross-organizational. %As such, the interaction in events is complex and intense, and most of the participants do not know each other well like in a company setting. 

\subsubsection{SLR}
There are no papers directly related to hybrid events in our SLR.

\subsubsection{Additional References}
As the hybrid event experiences are currently designed and being tested after the relaxation of COVID-19 restrictions around the world, research on hybrid events (as we understand and define them post-COVID-19) is evolving too. However, especially hybrid event experts from companies developing hybrid event technologies and organizers of hybrid events such as tourism and convention/conference agencies publish books/resources to support diverse stakeholders in moving their large events to virtual or hybrid level, and in this way creating flexible and safe experiences for attendants. Chodor's recent book ``Transitioning to Virtual and Hybrid Events: How to Create, Adapt, and Market an Engaging Online Experience'' is a relevant resource \cite{chodor2020transitioning}. 

\subsubsection{Future Relevance}
With the rising interest in hybrid events today, we believe that new forms of hybridity will emerge and interesting studies in different areas of research will follow. Benefiting from newly developed resources on hybrid events, not only from HCI and CSCW, but also from different disciplines, conducting real world studies on hybrid events, and developing and testing prototypes aiming at minimizing its problems and improving hybrid event experiences will be quite beneficial for both researchers of this phenomena and the hybrid event tourism market in making those events sustainable.

\subsection{Telepresence and Human-Robot Interaction}

Telepresence has been widely studied in HCI and CSCW and refers to the technologies enabling the (sense of) presence for a remote person. When telepresence is enabled through a robot, the remote person is also more active and capable of creating physical presence in a room.

\subsubsection{SLR}
Two full papers and one doctoral consortium paper related to telepresence and human-robot interaction are included in our SLR: \cite{rae_using_2013}, \cite{kim_is_2014}, \cite{rae_framework_2015}. Rae et al.'s paper from 2015 \cite{rae_framework_2015}, in particular, provides a rich categorization and framework of telepresence research. \autoref{tab:findings} presents their main findings where the Focal Point (FP) column tag is ``TP-Robot''. 

\subsubsection{Additional References}

The term telepresence may be used to refer to almost any system which enables HCM (e.g.~\cite{Riesenbach1994ontariotelepresenceproject, tanner2012sidebyside}), but its use is deeply enmeshed with \textit{robotic} presence at a distance (a.k.a. telexistence ~\cite{tachi_telexistence_2009}). We did not include telepresence as a term in our query because its literature is very large and diverse, even though its use is largely niche, and has a signficant amount of specialized findings that separate it from other HCM explorations. Arguably, some of this research on various issues of \textit{hybrid encounters} fits our interest in understanding hybridity, (as we will note below), however, a significant amount of the research also focuses on either engineering and/or human factors issues (e.g. methods of controlling robots~\cite{zhang2019handandgazecontrolofrobots}, or other exotic physicalized remote presence~\cite{leithinger2014physicaltelepresence}), and to a lesser extent the use of telepresence as a control mechanism in various contexts which either do not involve collaborators in the local activity space (e.g.~\cite{goza2004controlrobot}) or involves collaboration only with other remote users (e.g.~\cite{sabet2021telepresencedrones}. The large and diverse robotic telepresence literature would have overwhelmed the SLR's balance.

Robotic telepresence research has a lot to offer understandings of the potential for hybrid encounters, including studies comparing standard videoconferencing conditions to telepresence robot conditions (e.g. classroom learning (e.g.~\cite{tanaka2014telepresencerobot, shin2016telepresencerobot}), research on how and when remote participants seek help while using mobile robotic telepresence~\cite{boudouraki2021can}, the use of telepresence in otherwise collocated events (e.g. telepresence robots at conferences~\cite{rae2017telepresenceatscale}), crucial principles such as physical gestural mechanics of robots (e.g.~\cite{stahl2018gesturerobots}) or their effect on cultural expression (e.g.~\cite{shen2018robotsandculture}). Looking forward, some research is exploring telepresence robots with other AV modalities such as Mixed Reality~\cite{jones2021belonging}, adding complexity to improve the feeling of belonging on both sides of the connection.

\subsubsection{Future Relevance}

With the rise of mixed/virtual/augmented reality technologies and the COVID-19 reality, telepresence technologies and robots will be researched and developed more than ever, although the scalability of such technology will remain an issue.  

\section{Conclusion \& Future Directions}
\label{sec:conclusion}

%\section{Discussion}
%\label{sec:discussion}
%\textcolor{red}{Here we discuss what the most important findings tell us for the future of research. For example we could discuss what the learnings could be for the upcoming season of hybrid work as John Tang coined it.}

\sr{In this article, we provided a survey and a taxonomy on Hybrid Collaboration and Hybrid Meetings (HCM) research in the fields of HCI and CSCW by focusing on the publications in the ACM DL. Our literature review shows that HCM is still under-investigated by HCI researchers since even though 72 keywords were included in our search query, only 62 papers among 1,209 results fitted our criteria. Interestingly, even though tools supporting HCM have diversified and are widely available and used in many workplaces all around the world especially within the last decade, we observed a decline in the number of recent publications focusing on HCM in HCI and CSCW after 2014 and the lack of consensus in term usage among HCM researchers still exists. %since only one publication from 2017 \cite{xu_attention_2017} uses the term `Hybrid Meeting' and one publication from 2018 \cite{neumayr2018} uses the term `Hybrid Collaboration' explicitly in their studies. 
We also showed that HCM settings are quite diverse in these publications, for instance, the degrees of synchronicity and asynchronicity %and software tools and hardware devices used 
in the studies differ to a considerable extent. Yet, the number of locations are mostly limited to two and group sizes are mostly around ten or even less. We think that one motivation for such trends could be that most of these studies were aimed to be experimental. However, we already know that in the real modern workplace today -- even in the pre-COVID-19 world -- HCM is done in larger crowds with more co-located and remote ends \cite{saatcci2019hybrid}.}

%Lastly, 
One limitation in our methodology %might be due to the fact 
is that we used a keyword-based approach to select the publications initially (as opposed to a semantic approach). We followed the popular guidelines by Kitchenham and Charters \cite{kitchenham07} and were further inspired by recent applications of Nunes and Jannach \cite{nunes2017} as well as of Neumayr and Augstein \cite{neumayr2020}. Although we intended to be reasonably inclusive, still, some relevant keywords might have been missed and some related work is likely to have gone unnoticed. Maybe in the future, some reproducible alternative to the keyword-based approach for the selection of related publications in systematic literature reviews can be tried and tested. Yet, concerning the general trends and directions discussed in this article, and in connection to the answers to our research questions, we are optimistic that this limitation did not have a major effect. Another limitation of our article is that within the scope of this literature review we focused on the publications under the ACM DL only. We are aware of the fact that even before the pandemic, there were many HCM-focused articles published in different areas of research outside of HCI and CSCW, such as business communication, tourism, education, etc. The lack of consensus in term usage for describing HCM, and therefore lack of communication in related work also exist in other areas of research. A broader literature review covering these different areas of research would be necessary for researchers across diverse fields studying the same phenomenon to benefit from each other's research and communicate and collaborate with each other. %We hope our article can help researchers to get informed about the HCM research in HCI and CSCW in the pre-COVID-19 era and utilize our taxonomy to set and design their future research, and serves as an example for other systematic literature reviews on HCM in and across different disciplines.

\sr{%First and foremost, 
HCM have not received significant clear attention in HCI and CSCW, but this is, of course, now likely to change post-COVID-19. Of the research that does exist, much includes a hybrid setting without clearly exploring it, and most relies on controlled studies \tn{or lab experiments}. More real world studies are needed to better understand the particular dynamics of these meeting settings, not only to improve research, but because at the time of writing COVID-19 was still very much a pandemic -- and it will not be the last. In the case of COVID-19 %very practically in thebecause of the \tn{%Some thoughts: the time 
after governments in many countries %around the world 
eased curfew measures in the Summer and Fall of 2020 %after the initial outbreak of COVID-19 showed a first surge in 
hybrid scenarios surged.} %in many meetings as well as 

\sr{In education, %where 
hybrid teaching becomes the rule rather than the exception. Often education continues in the classrooms and lecture halls, while some students are quarantined due to positive test results of themselves or their contact persons. Others may have serious medical pre-existing conditions which makes it risky for them to attend class/lectures when the prevalence of an infectious disease is currently high. Especially in winter times, curfews may be imposed again, leading to more all-remote and less HCM. For diseases like COVID-19 which are thought to be correlated to seasonal changes depending on typical factors such as being outdoors in the sunlight, \textit{waves of increasing cases can be expected}. Therefore, the resurgence of HCM can be expected as we emerge out of the pandemic (or the next wave) in what has been called by John Tang a ``Season of Hybrid Work''. Experiences during the short 2020 season of hybrid work have shown many of the already known problems and solutions of both technical (e.g., a lack of awareness (mechanisms) \cite{xu_attention_2017}) and social nature (e.g., in-group/out-group effects \cite{bos_-group/out-group_2004,huang_preliminary_2006} or collocation blindness \cite{bos_collocation_2006}) that go along with this. We have to %make sure to 
be better prepared for the upcoming season(s) of hybrid work by learning from problems in the existing literature and following their suggestions on how to alleviate them. As it is hard for researchers to get an overview of existing literature due to the terminological confusion in this field, we %are confident 
hope that %our SLR and the taxonomy 
this review provides %can be 
a good starting point for finding the right links to matching use case %s concerning their 
domains and settings (e.g., group sizes, synchronicity \& asynchronicity, and tool \& device usage), and to learn from or to identify a design space for future research.}

\bibliographystyle{ACM-Reference-Format}
\bibliography{literature}

\appendix

\end{document}